\documentclass[preprint]{article}

\usepackage[utf8]{inputenc}
\usepackage[T1]{fontenc}
\usepackage{amsmath}
\usepackage{amssymb}
\usepackage{graphicx}
\usepackage{booktabs}
\usepackage{xcolor}
\usepackage{hyperref}
\usepackage{url}
\usepackage{microtype}
\usepackage{nicefrac}
\usepackage{subcaption}
\usepackage{natbib}

\usepackage{iclr2026_conference,times}
\iclrfinalcopy

\PassOptionsToPackage{dvipsnames}{xcolor}
\usepackage{hyperref}
\usepackage{url}
\usepackage{amsmath}
\usepackage{booktabs}
\usepackage{graphicx}   
\usepackage{multirow}   
\usepackage{float}       
\usepackage{array}       

\usepackage{microtype}
\usepackage{graphicx}
\usepackage{booktabs}
\usepackage[noabbrev,nameinlink]{cleveref}
\usepackage{listings}
\usepackage{makecell}
\usepackage{footnote}
\usepackage{dblfloatfix}
\usepackage[framemethod=TikZ]{mdframed}
\usepackage{multirow}
\usepackage{comment}
\usepackage{mathtools,tikz,caption}
\usepackage{subcaption}
\usepackage{adjustbox}
\usepackage[htt]{hyphenat}
\usepackage[frozencache,cachedir=minted-cache]{minted}
\usepackage{tcolorbox}
\usepackage{soul}
\usepackage{enumitem}
\usepackage{xfrac}
\usepackage{colortbl}
\usepackage[final]{pdfpages}

\usepackage{wrapfig}
\usepackage{mdframed}
\usepackage{lipsum}
\usepackage{tikz}
\usetikzlibrary{shapes.geometric}
\usetikzlibrary{tikzmark}
\usepackage[title]{appendix}
\usepackage{rotating}
\usepackage{adjustbox}

\usepackage[utf8]{inputenc} % allow utf-8 input
\usepackage[T1]{fontenc}    % use 8-bit T1 fonts
\usepackage{hyperref}       % hyperlinks
\usepackage{url}            % simple URL typesetting
\usepackage{booktabs}       % professional-quality tables
\usepackage{amsfonts}       % blackboard math symbols
\usepackage{nicefrac}       % compact symbols for 1/2, etc.
\usepackage{microtype}      % microtypography
\usepackage{xcolor}         % colors
\definecolor{color1}{rgb}{0.1216,0.4667,0.7059}
\definecolor{color2}{rgb}{1.0,0.498,0.0549}
\definecolor{color3}{rgb}{0.1725,0.6275,0.1725}
\definecolor{color4}{rgb}{0.8392,0.1529,0.1569}
\definecolor{color5}{rgb}{0.5804,0.4039,0.7412}
\definecolor{color6}{rgb}{0.549,0.3373,0.2941}
\definecolor{color7}{rgb}{0.8902,0.4667,0.7608}
\definecolor{colorXXX}{rgb}{0.498,0.498,0.498}
\definecolor{color8}{rgb}{0.7373,0.7412,0.1333}
\definecolor{color9}{rgb}{0.0902,0.7451,0.8118}

\newcommand{\gail}[1]{\textcolor{color3}{[Gail]: #1}}

\newcommand{\ray}[1]{\textcolor{color5}{[Baishakhi]: #1}}

\newcommand{\wei}[1]{\textcolor{red}{[Wei]: #1}}

{\end{mdframed}
\vspace{2pt}
\end{minipage}}

\newcommand{\bench}{{CodeSense }}

\newcommand{\TODO}[1]{\textbf{\color{red}TODO:{ #1} }}

\newcommand\figref[1]{Fig.~\ref{#1}}

\newcommand{\fakeparagraph}[1]{\noindent\textbf{#1.}}

\definecolor{junglegreen}{rgb}{0.16, 0.67, 0.53}

\definecolor{ddw}{cmyk}{0, 0.7808, 0.4429, 0.1412}

  {}             % End block - do nothing

  {\noindent\begin{flushleft}\small\ttfamily}  % Begin block
  {\end{flushleft}}                  % End block

\newcommand{\mycomment}[1]{}

\definecolor{mygray}{RGB}{53,56,57}
\DeclareRobustCommand\sampleline[1]{%
  \tikz\draw[#1,dashed,mygray,line width=1.3pt,dash pattern=on 3pt off 1pt] (0,0) (0,\the\dimexpr\fontdimen22\textfont2\relax)
  -- (1.7em,\the\dimexpr\fontdimen22\textfont2\relax);%
}
\definecolor{Periwinkle}{rgb}{0.8, 0.8, 1.0}

\usepackage{wasysym}  

\title{CodeSense: a Real-World Benchmark and Dataset for Code Semantic Reasoning}

\author{
Monoshi Kumar Roy\textsuperscript{1}
\And
Simin Chen\textsuperscript{2}
\And
Benjamin Steenhoek\textsuperscript{3}
\And
Jinjun Peng\textsuperscript{2}
\And
Gail Kaiser\textsuperscript{2}
\And
Baishakhi Ray\textsuperscript{2}
\And
Wei Le\textsuperscript{1}
\\[1ex] 
\AND
\textsuperscript{1} \text{Iowa State University}\\\texttt{\{monoshi,weile\}@iastate.edu}
\And
\textsuperscript{2} \text{Columbia University}\\\texttt{\{sc5687,jinjun,kaiser,rayb\}@cs.columbia.edu}
\And
\textsuperscript{3} \text{Microsoft Data \& AI}\\\texttt{bensteenhoek@microsoft.com}
}

\begin{document}

\maketitle

\begin{abstract}
Understanding and reasoning about code semantics is essential for enhancing code LLMs' abilities to solve real-world software engineering (SE) tasks. Although several code reasoning benchmarks exist, most rely on synthetic datasets or educational coding problems and focus on coarse-grained reasoning tasks such as input/output prediction, limiting their effectiveness in evaluating LLMs in practical SE contexts. To bridge this gap, we propose \bench, the first benchmark that makes available a spectrum of fine-grained code reasoning tasks concerned with the software engineering of real-world code. We collected Python, C and Java software projects from real-world repositories. We executed tests from these repositories, collected their execution traces, and constructed a ground truth dataset for fine-grained semantic reasoning tasks.  We then performed comprehensive evaluations on state-of-the-art LLMs. Our results show a clear performance gap for the models to handle fine-grained reasoning tasks. Although prompting techniques such as chain-of-thought and in-context learning helped, the lack of code semantics in LLMs fundamentally limits models' capabilities of code reasoning. Besides dataset, benchmark and evaluation, our work produced an execution tracing framework and tool set that make it easy to collect ground truth for fine-grained SE reasoning tasks, offering a strong basis for future benchmark construction and model post training. Our code and data are located at \url{https://codesense-bench.github.io/}.

\end{abstract}
%\gail{ I realize this is mostly notes at this point anyway, but this abstract is much too long.  The abstract should be short enough to leave enough space for the Introduction heading and the Intro's first paragraph to fit on the first page. }

\section{Introduction}
\label{sec:intro}
%\wei{Ray and I both had a pass on intro, please edit important changes using colored fonts}\gail{I put in review mode, hope this works. ... Finished intro. }

Semantic code reasoning—the capacity to understand and predict the behaviour of software—is a core requirement underpinning a wide range of complex software engineering (SE) tasks, including test input generation, vulnerability detection, fault localization, bug repair, refactoring, and functional verification. Unlike syntactic pattern matching, which may rely on token-level similarity or statistical regularities, semantic reasoning ("codesense") entails a deep, execution-oriented understanding of how software operates. Although code semantics can be expressed in many ways, in practice, developers engage in semantic reasoning through tasks like predicting a function’s input-output behavior, tracing variable values, analyzing control flow paths, identifying loop invariants, etc. This form of reasoning aligns with formal definitions from programming language theory—particularly operational semantics, which models step-by-step execution, and axiomatic semantics, which uses logical assertions to describe program properties. Such reasoning tasks also reflect the real-world demands placed on developers and provide a natural grounding for their day-to-day work.

Recent years have witnessed the emergence of numerous benchmarks for evaluating coding-related tasks. However, the majority of these efforts have focused on code generation using synthetic or narrowly scoped data—for example, HumanEval+ ~\citep{liu2023codegeneratedchatgptreally}, LiveCodeBenchmark~\citep{jain2024livecodebenchholisticcontaminationfree}, Bigcodebench~\citep{zhuo2024bigcodebenchbenchmarkingcodegeneration}, and CodeBenchGen~\citep{xie2024codebenchgencreatingscalableexecutionbased}—often extracted from isolated competitive programming problems. Consequently, they fail to capture the complexity and structure of real-world software development. Other benchmarks that incorporate real-world code, such as SWE-Bench~\citep{jimenez2024swebenchlanguagemodelsresolve}, SWE-PolyBench~\citep{rashid2025swepolybenchmultilanguagebenchmarkrepository}, and KGym~\citep{mathai2024kgym}, tend to evaluate only task-specific performance (e.g., patch generation for GitHub issues), making it difficult to assess whether models exhibit generalizable semantic understanding. Finally, reasoning-focused benchmarks, such as CruxEval~\citep{gu2024cruxevalbenchmarkcodereasoning}, primarily target function-level input/output prediction over short and synthetic code fragments involving random string operations. Such settings neglect the fine-grained semantic reasoning about internal program behavior and properties, data dependencies, and control structures required to solve a variety of SE tasks for complex real-world software systems.

To this end, we propose CodeSense, a benchmark for fine-grained code semantic reasoning, constructed from real-world GitHub projects in Python, C, and Java (see Table~\ref{tab:related}). \bench~introduces a spectrum of reasoning tasks at statement, code-block, and function levels, targeting essential semantic properties frequently needed across SE activities.  For example, predicting loop iteration counts is critical for input/output prediction, performance analysis, and detecting infinite loops (e.g., denial-of-service vulnerabilities). Branch condition prediction and reasoning about pointers in C code are important for test input generation and memory safety assurance. As illustrated in Figures~\ref{fig:test} and~\ref{fig:vul}, fine-grained semantic reasoning about arithmetic operations, control flow, and API semantics is foundational for a variety of SE applications. Prior work~\cite{ding2023tracedexecutionawarepretrainingsource,ding2024semcodertrainingcodelanguage} has shown that incorporating semantic signals during training improves model performance on code generation, branch prediction, code clone detection, program repair and vulnerability detection tasks, motivating our design of \bench to comprehensively evaluate models' capabilities for semantic reasoning.

\begin{figure}
% \begin{wrapfigure}{r}{0.65\textwidth}
    \centering
    \begin{subfigure}[t]{0.47\textwidth}
    \centering
    \begin{tcolorbox}
    \begin{minted}
    [
    frame=lines,
    numbersep=2mm,
    breaklines=true,
    fontsize=\tiny,
    highlightlines={3},
    highlightcolor=yellow,
    linenos,
    tabsize=2,
    breaklines,
    ]
    {diff}
    void foo(int input){
      int n = input * 23;
      if (3465>=n>=2287){
         //dangerous code need to be tested 
      }
    }
    \end{minted}
    \end{tcolorbox}
    \caption{Test input generation and program execution reasoning: to generate a test input that can lead to the execution of dangerous code at line~4, the model needs to find an input that can satisfy the branch condition {\tt 3465$\geq$n$\geq$2287at line~3}. Understanding the semantics of operators '*' at line 2 and '$\geq$' at line 3 is needed to effectively generate an input that reaches the dangerous code, e.g., input = 120 \label{fig:test}. The branches and arithmetic operations can be quite diverse in different programs, and it is hard to generalize patterns regarding which code text should use what kind of test input to execute.}
    
  %  However, it is hard to generalize across examples, as each program can have different values and, without knowing semantics, it is hard to find patterns. 
    \end{subfigure}\quad%
    \begin{subfigure}[t]{0.5\textwidth}
    \centering
    \begin{tcolorbox}
    \begin{minted}
    [
    frame=lines,
    numbersep=2mm,
    breaklines=true,
    fontsize=\tiny,
    highlightlines={3,5,7,8},
    highlightcolor=yellow,
    linenos,
    tabsize=2,
    breaklines,
    ]
    {diff}
    void bar (int nbits) {
        FFTContext *s = malloc(sizeof (* s));
        if (s && nbits==100))
            free(& s);
        else ... return s;
 }
    \end{minted}
    \end{tcolorbox}
    \caption{Vulnerability detection, fault localization and program repair: this code has a memory leak vulnerability when the {\tt if} condition at line~3 is {\tt false}. To detect this vulnerability, the model should know that calls to {\tt malloc} and {\tt free} are related to the vulnerability (semantics of the API calls), and should be paired along the program paths along both branches starting at line~3 (semantics of the branch statements and control flow). Similarly, {\tt malloc} and {\tt free} can be located in various code contexts in different programs, and thus it is hard to generalize the patterns only from the code text, without semantic code reasoning.
    \label{fig:vul}}
    \end{subfigure}
    \definecolor{DiffGreen}{HTML}{368B00}
    \definecolor{DiffRed}{HTML}{A51901}
    \caption{Fine-grained code semantics are the keys for solving many SE tasks
    \label{fig:example}}
\end{figure}
%\wei{Wei: proofread the text in figures}

%implementation

\begin{table}[htbp]
  \centering
  \caption{Optimal design space of code reasoning benchmarks (\Circle \space denotes not support, \LEFTcircle \space denotes partial support, and \CIRCLE \space denotes fully support)}
  \resizebox{\textwidth}{!}{
    \begin{tabular}{lcccccccc}
    \toprule
    \toprule
    \textbf{Benchmark} & \textbf{Real-World Projects} & \textbf{Multi-lingual} & \textbf{Function I/O} & \textbf{Fine-Grained Reasoning} & \textbf{Exec. Steps} & \textbf{API Understanding} & \textbf{Multi-File Context} & \textbf{Realistic Project Structure} \\
    \midrule
    \textbf{CruxEval \citep{gu2024cruxevalbenchmarkcodereasoning}} & \Circle & \Circle & \CIRCLE & \Circle & \Circle & \Circle & \Circle & \Circle \\
    \textbf{CruxEval-X \citep{xu2024cruxeval}} & \Circle & \CIRCLE & \CIRCLE & \Circle & \Circle & \Circle & \Circle & \Circle \\
    \textbf{REval \cite{chen2024reasoningruntimebehaviorprogram}} & \Circle & \Circle & \CIRCLE & \LEFTcircle & \LEFTcircle & \Circle & \Circle & \Circle \\
    \textbf{CodeMind \citep{liu2024codemindframeworkchallengelarge}} & \CIRCLE & \CIRCLE & \CIRCLE & \Circle & \Circle & \CIRCLE & \CIRCLE & \CIRCLE \\
    \textbf{CoRe \citep{xie2025corebenchmarkingllmscode}} & \Circle & \CIRCLE & \Circle & \CIRCLE & \Circle & \Circle & \Circle & \Circle \\
    \midrule
    \textbf{CodeSense} & \CIRCLE & \CIRCLE & \CIRCLE & \CIRCLE & \CIRCLE & \CIRCLE & \CIRCLE & \CIRCLE \\
    \bottomrule
    \bottomrule
    \end{tabular}%
  }
  \label{tab:related}%
\end{table}%

 To construct \bench, we collected 544 Python, 100 C, and 100 Java projects.  To enable supervised evaluation, we develop a framework that automatically extracts ground-truth annotations for these tasks using static and dynamic program analysis.  We build, execute and log run-time execution values and traces of the real-world projects. Our dataset includes a total of 2125 Python,  876 C, and 875 Java unique functions, based on which, we curated 4495 samples with their ground truth in our benchmark. Please see more details in Section 2.
% \gail{say something very brief about what we mean by filtering here, e.g., "After filtering for xxx, we obtained ..."}
%For Python, we executed test cases using the pytest module, and collected execution traces using PySnooper. For C, we used the projects curated in OSS-Fuzz, and the projects were built and fuzzed by OSS-fuzz's docker-based infrastructure. We employed GNU debugger(GDB) to collect run-time traces from C projects. For Java, We used the SF110 dataset, and EvoSuite to generate test cases. We developed the Java tracing framework using the Java Debugger. Using the collected execution trace logs, we build analysis tools to extract the ground truth for our reasoning tasks, and applied filters (details in Section 2) and selected a total of 2125 Python,  876 C, and 875 Java unique functions in our benchmark.

%we constructed \bench, by extracting the function-level traces and curating benchmark datasets for different code understanding tasks.

%\wei{Roy please add some details regarding our benchmark construction and tools} 

%research question
Using our benchmark, we evaluated 14 state-of-the-art (SOTA) LLMs and investigated six research questions regarding the models' code semantics reasoning capabilities. Previous work has shown that models did not perform well on code reasoning tasks such as input/output prediction~\cite{gu2024cruxevalbenchmarkcodereasoning} and vulnerability detection~\cite{steenhoek2025errmachinevulnerabilitydetection}; to understand why models fail and identify places for improvement, we investigate:
\textbf{RQ1:} Does increasing code size make semantic reasoning more difficult?  
\textbf{RQ2:} Which types of program statements are easier or harder for models to reason about? % in value prediction?  
\textbf{RQ3:} How do models perform on code properties critical for SE tasks, such as predicting pointer aliasing, loop iteration counts, and branch conditions?  
\textbf{RQ4:} How effective are different prompting strategies in improving semantic reasoning?   
\textbf{RQ5:} Can models reason approximately when exact values or semantics are hard to infer?
\textbf{RQ6:} Do models perform better on some programming languages than others?

% {\bf RQ1}: {\it What types of program statements the model can understand well?  does the model feel any type particularly difficult?}  and {\bf RQ2:} {\it Does increasing the size of code increase the difficulty of code reasoning?}. We identified and evaluated frequently code reasoning tasks shared across many SE tasks, and investigated {\bf RQ3}: {\it How do models perform regarding code properties that are frequently needed for solving SE tasks such as pointer aliasing, loop iterations and values, and branch prediction?} We also studied {\bf RQ4}: {\it How do different prompt techniques help?} and {\bf RQ5}: {\it Any particular programming languages are easier for the models?}. We observed that models face challenges of predicting a precise program value/property, so  we investigated {\bf RQ6}: {\it Can models reason about an approximation of code semantics?}.
%%While the code reasoning capabilities of LLMs rapidly improve, to probe we are there yet, we studied {\bf RQ7}: {\it How does the best SOTA LLMs perform our fine-grained code reasoning tasks?}

%results
%\wei{Roy: can you add a summary of results}\\
Our results reveal that current LLMs, including SOTA models like Claude 3.5 ~\cite{anthropic2024claude35}, GPT-4o-mini ~\cite{openai2024gpt4omini}, and Gemini 1.5 ~\cite{google2025gemini15flash} struggle with fine-grained code semantics. They often fail to reason about even single statements from real-world code—particularly arithmetic expressions and API calls—and perform poorly on tasks involving loop values and iteration counts. Basic chain-of-thought prompting offers limited benefit, and few-shot prompting yields only modest improvements. In-context learning is most effective when prompts define new concepts or include highly relevant examples.  Interestingly, models can correlate natural language code semantics questions with certain code patterns. For instance, when the code contains assignments like {\tt p = q},  models correctly respond to the prompt  "do {\tt p} and {\tt q} at <line> alias the same memory address" even in zero-shot settings. Similarly, models reliably infer loop bounds in explicit cases such as {\tt for i in range(100):}. Among the 14 models evaluated, Claude 3.5 consistently achieved the best performance. We also observe that Java and Python code are generally easier for models to reason about than C, and that input prediction (i.e., reverse semantic inference) remains among the most challenging tasks.

\noindent\textbf{Contributions.} This work introduces CodeSense, a realistic and comprehensive benchmark for evaluating LLMs' fine-grained code semantics reasoning in practical software engineering contexts. We advance the state-of-the-art code reasoning benchmarks by:

\begin{enumerate}[topsep=0pt,itemsep=-1ex,partopsep=1ex,parsep=1ex]
\item Defining a diverse set of fine-grained semantic reasoning tasks grounded in real-world software engineering needs,
\item Developing a scalable open-source framework and toolchain to automatically generate execution traces and semantic annotations, enabling continuous benchmark expansion while mitigating data leakage,
\item Constructing a benchmark dataset using real-world projects in Python, C, and Java,
\item Empirically analyzing six research questions across 14 state-of-the-art LLMs to assess their strengths and limitations in semantic reasoning, and 
\item Launching a public leaderboard to support reproducibility and accelerate progress on semantic reasoning for code: \url{https://codesense-bench.github.io/leaderboard.html}
\end{enumerate}

\mycomment{

\TODO{LLM is used for SE applications}
Recently, code large language models (LLMs) have demonstrated great potential in software development tasks such as code generation, testing, and maintenance. These models are trained on massive corpora of both natural and programming languages and are typically evaluated using benchmarks that measure their performance on specific tasks. For instance, HumanEval and DyCodeEval have been proposed to benchmark code generation capabilities. 

\TODO{existing benchmarks do not consider code semantics but understanding code semantics is mandatory for many SE tasks, such as...}
While these existing benchmarks are valuable, they primarily focus on task-specific evaluation rather than assessing a model’s code semantic understanding-—a critical capability for enabling automatic reasoning about program properties. In traditional software development, developers first understand the semantics of the code and then write software accordingly. In contrast, code LLMs often generate code directly, raising an important question: \textit{Do they truly understand the semantics of the code?}
Semantic understanding is especially important for tasks such as test generation and vulnerability detection. As illustrated in \figref{fig:vulnerability-examples}, \TODO{explain the examples}, if code LLMs can accurately comprehend the semantics of the highlighted code snippets, they are more likely to produce correct, secure, and testable code. Enhancing semantic understanding in LLMs could therefore lead to more reliable and effective tools for real-world software engineering.

\TODO{existing work and their limitation}
To explore whether code LLMs are capable of understanding code semantics, a number of code reasoning benchmarks have been proposed. Unlike benchmarks that evaluate real-world task-solving capabilities, these reasoning benchmarks focus on code semantic understanding. For example, \TODO{}. Unfortunately, current code semantics benchmarks have several key limitations that prevent a comprehensive understanding of code LLMs' semantic capabilities:  \textit{1. Non–real-world code sources}: Existing benchmarks are often constructed from synthetic datasets or coding contest problems instead of real-world software repositories. This disconnect limits the benchmarks’ ability to reflect the kinds of semantic patterns and challenges encountered in real-world software engineering. \textit{2. Coarse-grained semantic focus}: Many benchmarks emphasize coarse-grained semantics, typically at the function level. However, coarse granularity fails to capture subtle, fine-grained semantic nuances necessary for reasoning about edge cases. This can be particularly problematic in security-critical scenarios like vulnerability detection, where small semantic misunderstandings can lead to significant failures. \textit{3. Toy semantic tasks}: Existing benchmarks often focus on simplified tasks like input/output prediction or basic code comprehension, which are far removed from the complex reasoning required in real-world SE. As a result, they provide limited insight into a model’s ability to understand and reason about intricate program behaviors.

\TODO{our proposal and highlevel solution} 
To address the limitations of existing work and enable fine-grained semantic understanding in real-world software projects, we propose \tool. Unlike prior benchmarks based on synthetic or educational data, \tool collects its codebase directly from real-world GitHub projects, thereby addressing the first limitation related to unrealistic data sources. To overcome the second limitation—-coarse-grained semantic evaluation—-\tool introduces a diverse set of semantic understanding tasks across multiple levels of granularity, addressing statement-level, block-level, and function-level reasoning. This allows for a more detailed and layered evaluation of a model's semantic capabilities. In addition, \tool incorporates two critical code property reasoning tasks: loop analysis and pointer analysis. These tasks are essential and widely applicable in real-world SE. For example, loop analysis can be used to reason about software performance, while pointer analysis plays a key role in detecting memory safety issues and vulnerabilities. Together, these design choices allow \tool to provide a more realistic and comprehensive benchmark for evaluating the semantic understanding of code LLMs in practical, real-world contexts.

\TODO{implementation and evaluation}

\TODO{contribution}
Our primary contributions are:
\begin{enumerate}

% \item We defined a spectrum of fine-grained code reasoning tasks for evaluating and improving models' understanding of code semantics, which we show are directly linked to handling SE tasks such as test input generation, vulnerability detection, debugging, program repair and code generation. 

% \item We constructed a benchmark and dataset, consisting of real-world software projects in multiple programming languages of C,  python, Java and Javascripts 

% \item We developed frameworks and tools that enable automatic collection of fine-grained code execution data for future generation of data and bechnmarks. The data including program input/output, intermediate values, traces and program properties like loop counters. This framework can be used to generate further program execution data an benchmarks.

% \item We performed systematic evaluate on the models' capabilities of understanding code semantics and present interesting findings.
\end{enumerate}

\gail{can we say anything at the end of the introduction about a leaderboard?}

\clearpage

\section{Introduction} 
(1) code semantics is important for many SE engineering tasks, which models do not perform very well (Test input generation, debugging, repair we need to understand the semantics of code? are important for those tasks? Edit and repair, partial 

(2) Does the model understand the semantics code?

(3) in the past, the evaluation is coarsed-grained, only limited to input/output, but for the variety applications, we need to know other code reasoning reaability

\wei{Todo: need to add 4 examples to show the importance}

~\cite{semcoder} Code semantics help code generations 

\begin{figure}
% \begin{wrapfigure}{r}{0.65\textwidth}
    \centering
    \begin{subfigure}[t]{0.45\textwidth}
    \centering
    \begin{tcolorbox}
    \begin{minted}
    [
    frame=lines,
    numbersep=2mm,
    breaklines=true,
    fontsize=\tiny,
    highlightlines={3},
    highlightcolor=yellow,
    linenos,
    tabsize=2,
    breaklines,
    ]
    {diff}
void foo(int input){
   int n = input * 23;
  if (n >= 1580){
     //dangerous code need to be tested 
  }
}
    \end{minted}
    \end{tcolorbox}
    \caption{Test input generation and execution reasoning: to generate a test input that can lead to the execution of dangerous code. The model needs to find an input that can make the condition $n\geq$ 1580 true. However, it is hard to generalize across examples, as each program can have different values, without knowing semantics, it is hard to find patterns. Understanding the semantics of 'x' at line 2 an d'$\geq$' at line 3 is needed to effectively generate test input that reaches dangerous code, e.g., n = 2000 \label{fig:vulnerability-example-bof}}
    \end{subfigure}\quad%
    \begin{subfigure}[t]{0.5\textwidth}
    \centering
    \begin{tcolorbox}
    \begin{minted}
    [
    frame=lines,
    numbersep=2mm,
    breaklines=true,
    fontsize=\tiny,
    highlightlines={3,5,7,8},
    highlightcolor=yellow,
    linenos,
    tabsize=2,
    breaklines,
    ]
    {diff}

    void foo (int nbits) {
        FFTContext *s = malloc( sizeof (* s));
        if (s && init (s , nbits)){
            if (s==p) nbits = 100;
            free(& s);
        }
         return s;
 }
    \end{minted}
    \end{tcolorbox}
    \caption{Vulnerability detection, fault localization and repair: This code has a memory leak vulnerability when the if condition at line~4 is false. To detect this vulnerability, models need to know malloc and free should be paired along the program paths, but here when if is false at line~4, the memory leak can happen. It is hard to generalize the patterns beause there can be many ifs before free.
    \label{fig:vulnerability-example-npd}}
    \end{subfigure}
    \definecolor{DiffGreen}{HTML}{368B00}
    \definecolor{DiffRed}{HTML}{A51901}
    \caption{Examples of vulnerability detection as a complex code reasoning task. Diffed lines (\textbf{{\color{DiffGreen}+}}/\textbf{{\color{DiffRed}-}}) show the lines changed to patch the vulnerability.

    \label{fig:vulnerability-examples}}
\end{figure}

debugging discusss with the second example

\wei{We studied the following research questions:}
Motivation: Perform bad on code reasoning tasks of input/output prediction, we want to understand why models fail and what are the challenges of the models. So defined fine-grain tasks and research questions (RQ1 and RQ2). In our benchmark, we also highlight the frequent tasks  that are important for many SE tasks (RQ3) and understand do simple prompt can help (RQ4), and what types of programming languages need more help (RQ5).

{\bf RQ1}: What types of program statements the model can understand well?  does the model feel any type particularly difficult? 

{\bf RQ2:} Does increasing the size of code increase the difficulty of code reasoning?

Two experiments 
input output prediction for small, medium and large code 

{\bf RQ3}: How do model perform regarding queries that are frequently needed for solving SE tasks such as properties related to loops and pointers?

{\bf RQ4}: How do different prompt techniques perform?

{\bf RQ5}: Any particular programming languages are easier for the models?

\wei{Contributions}
This paper presents a comprehensive study and real-world multi-lingual benchmark, dataset and framework tools that can evaluate the future models. It's the first this paper is first that defines the code reasoning capabilities via a spectum of fine-grained code reasoning tasks that directly link to important SE applications. In summary, we make the following contributions:
\begin{enumerate}
\item We defined a spectrum of fine-grained code reasoning tasks for evaluating and improving models' understanding of code semantics, which we show are directly linked to handling SE tasks such as test input generation, vulnerability detection, debugging, program repair and code generation. 
\item We constructed a benchmark and dataset, consisting of real-world software projects in multiple programming languages of C,  python, Java and Javascripts 
\item We developed frameworks and tools that enable automatic collection of fine-grained code execution data for future generation of data and bechnmarks. The data including program input/output, intermediate values, traces and program properties like loop counters. This framework can be used to generate further program execution data an benchmarks.
\item We performed systematic evaluate on the models' capabilities of understanding code semantics and present interesting findings.
\end{enumerate}

}
\section{Benchmark Construction}
\label{Benchmark}

\subsection{Defining a Spectrum of Code Reasoning Tasks}
To design tasks for evaluating LLMs' capabilities of code semantic reasoning, we first considered the definition of code semantics. In programming languages and software engineering, code semantics ---``what is the meaning of this code'' --- are defined as what is the output value given the input of a code snippet. Such fine-grained reasoning tasks are directly related to end-tasks in software engineering. For example, previous work~\citep{ding2023tracedexecutionawarepretrainingsource} shows that when fine-tuned with statement-level values, the performance of the models improved for vulnerability detection, branch prediction and code clone detection. Prior study~\citep{steenhoek2025errmachinevulnerabilitydetection} reported that although recent LLMs improved math reasoning and natural language reasoning significantly, they are still insufficient for handling end-tasks related to code reasoning. To help locate the weakness of models' code reasoning at a fine-granularity and help models to improve a variety of SE applications that are linked to the fine-grained reasoning steps, we designed the following code reasoning tasks:

\begin{figure}[t]
    \centering
    \includegraphics[width=0.82\linewidth,height=0.7\textheight,keepaspectratio]{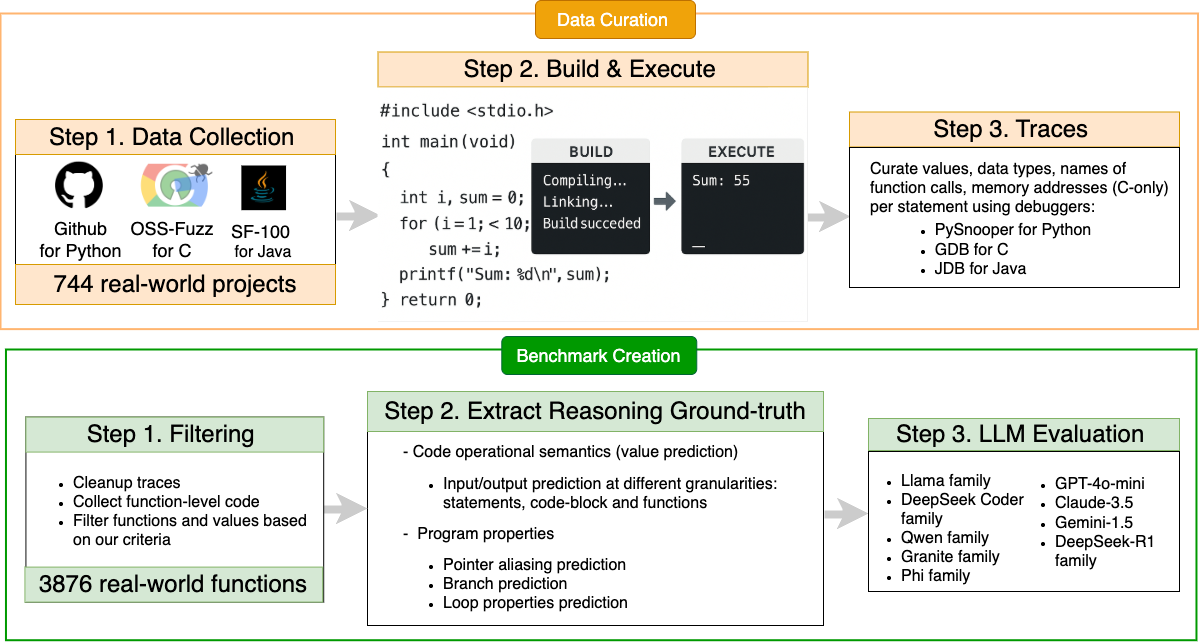}

    \caption{CodeSense: Data curation and benchmark creation overview.}
    \label{fig:overview}
\end{figure}

%\wei{Overall, we have defined tasks that evaluate operational semantics of code; that is, predicting values at statement, code-block and function level. We also defined tasks related to axiomatic semantics of code; that is predicting program properties for various code structures, including loops, branches and pointers.}
 
 %compared to previous benchmarks based on end-user SE tasks such as predicting input/output of a function~\citep{xxx} and generating patches to resolve GitHub issues,

\noindent {\bf Task 1: Block-level code semantics (RQ1)}: To investigate whether a model understand a chunck of code, we give a block of statements. We give input and ask the models to predict the execution output; also we give output, we ask the models to predict the input. Input/output prediction of a function is a special case of probing block-level code semantics. In our evaluation, we sampled a block of statements from the entry of the functions and increased the sizes of blocks, including the entire function.

\noindent {\bf Task 2: Statement-level code semantics (RQ2)}: We classified program statements based on programming language semantics and evaluated models on five common statement types, including {\it arithmetic}, {\it boolean expression}, {\it API/Function call}, {\it variable assignment} and {\it constant assignment}. Knowing code semantics at the statement level means that given an input of the statement, the models are able to produce a correct output. In our evaluation, we studied output predictions in more depth. We randomly sampled a statement from the program and asked the models to predict its output given input. %(see the prompt in Appendix/data package).~\ray{ano input prediction?}

%we prompted the model what would be the value of the statement execution. We discarded comments, blank lines, and statements that are part of control flow structure, if-else conditions, loops (for, while), and try-except blocks, and sampled statements from varying block sizes of 1 to 10. 

\noindent {\bf Task 3: Code properties within a function (RQ3)}: Code property (property regarding a particular code construct) is another aspect of code semantics. We focused on three important properties in this benchmark. Loops are related to code optimization and detecting bugs.  Reasoning about pointers in C code are very important for assuring memory safety and detecting and repairing vulnerabilities. Knowing how to predict branch outcomes can help generate test inputs and parallelize code.

%we asked models to predict loop (what are the loop iterations and values in and after the loop), pointers (whether two pointers are aliasing), branch (whether the true/false branch will be executed).

\noindent {\bf Task 3-1: Loop property.} Given an input of a function, we asked models to predict the number of loop iterations, the values in the loop and the values after executing the loop. In our evaluation, we randomly sampled a loop in a function and randomly sampled variables in and after the loop.

\noindent {\bf Task 3-2: Pointer property.} Here, we give the models a function and its input, and we ask models to predict whether the two pointers are aliased (pointing to the same memory location) at a given program point. 

\noindent {\bf Task 3-3: Branch property.} Given a function and its input as well as the location of a conditional branch in the function, we ask the model to predict what is the outcome of the branch. In our evaluation, we randomly selected a conditional branch in the function for prediction.

\noindent {\bf Task 4: Approximation of code semantics (RQ5)}: Reasoning about concrete values for the above tasks is very challenging. Sometimes, to solve an SE task, we may only need an approximate value of code semantics. For example, in \figref{fig:test}, the models do not have to generate a concrete number like {\tt input=120}; it is sufficient for models to tell us that an integer input between 100-150 can trigger the dangerous code. We designed a set of {\it abstract values} for different data types, following prior literature~\citep{ding2023tracedexecutionawarepretrainingsource} and evaluate if the models can predict abstract values correctly.

%{\bf RQ1}: (statement-level) What types of program statements can the model understand well?  does the model feel any type particularly difficult? 

%\roy{Table 1 in ~\citep{steenhoek2025errmachinevulnerabilitydetection} shows that models can have great math reasoning capabilities, but still are weak for handling code semantic reasoning tasks. In \figref{fig:vul} in our paper, we explained how knowing values is related to predicting vulnerabilities.}

The above tasks are also used for studying {\bf RQ4} regarding prompting techniques and {\bf RQ6} comparing different programming languages. We have included the prompts for all the above tasks in Appendix/data package.

%\TODO{Table for each type of statement}

%Presentation: slide 1: concrete vs quantized value comparison, average overall statements: 1 figure slide 2: individual statements comparison (average overall all models, individual models) 4 figures \wei{[done]}

%{\bf RQ2:} (block) Does increasing the size of code increase the difficulty of code reasoning?

%{\bf Task2}: We ask model to predict the output of a chuck statement given output, Input/output prediction for 1 line, 2, ... 10 lines of code, entire function 
\mycomment{
Presentation: 
Plot 1: input/output prediction for block size 1-3, both quantized and concrete 
Plot 2: input output prediction for small, medium and large code (lines of code), overall performance
both concrete and quantized

{\bf RQ3}: (Query) How do model perform regarding code properties that are frequently needed for solving SE tasks?

Sampling based 
{\bf Task 3}: alias pointers
{\bf Task 4}: branch prediction
{\bf Task 5}: loop prediction

count, in-loop value, after loop value (concrete/quantized)

{\bf RQ4}: (prompt) How do different prompt techniques perform? \wei{mostly done}

Presentation: 
[done] Plot 1:  0-3 shot plot for all the models  (does add shots help)
[done] Plot 2: 3-shot general vs same function for all the models (does same function/RAG help)

Plot 3: best above + COT (model specific)  (does COT help)

for all the tasks Statement, block, query prediction: Quantized and concrete 

{\bf RQ5}: (language) Any particular programming languages are easier for the models?

Input/output prediction for multiple languages: java

paid models 
}

\subsection{Collecting and Tracing Real-world Multi-Lingual Software Projects}
%\wei{Roy: separate the current giant figures into left samll component based figure and a table on the right}
Constructing ground truth for the set of code-semantics tasks is a great challenge. 
We collected a total of 744 real-world software projects of Python, C and Java from GitHub. We developed a framework and tool chain to build the projects, run tests and collect the execution traces which contain values, data types, names of function calls and memory addresses (for C code) at each statement. We developed analysis tools to extract ground truth for the benchmark tasks from those fine-grained code semantics data. Please check our Appendix ~\ref{app:language_rationale} for our language selection rationale and how our framework can be easily extended to other languages.

%We gathered open-source programs from the wild, which are in real-world use. Finding a diverse set of programs that we can compile/run and input into these programs becomes a challenge. We collected programs using the most efficient resources available.
%and to ensure we don't miss out with the recent projects, we used Github API to search Python Projects based on the number of stars.

\fakeparagraph{Python} We collected 1489 GitHub repositories from the PyPIbugs dataset~\citep{allamanis2021self}. We removed projects that don't contain test cases or have not been updated in the last four years, and obtained 544 projects.  We first installed dependencies for each project and used {\tt pytest}~\citep{pytest} to run tests, and {\tt Pysnooper} ~\citep{rachum2019pysnooper} for tracing. % We recorded variable values and types for each statement, excluding the traces in the built-in functions or third-party dependencies. \\

%We used the {\tt pytest} module for testing. 

%and we made sure the projects that we selected have a testing folder in their repo. For further filtering, we discarded the projects that had not been updated in the last four years and for python we selected 544 projects. We first installed the common dependencies in the independent environment for each project, executed the test cases in the projects using pytest. For tracing, we used Pysnooper, keeping track of the source code files in the project directory, excluding the traces in the buiit-in functions or third-party dependencies, and for variables, we kept track of both their values and types. \\

\fakeparagraph{C} We used 100 projects curated in OSS-Fuzz~\citep{ossfuzz}. We built and fuzzed the real-world projects using the OSS-Fuzz infrastructure in the docker environment with project-wise fuzzing harnesses. We developed a tracing framework built on the GNU debugger (GDB) ~\citep{gdb}. %The framework records function calls, variable values, and other program states during run-time. 

\fakeparagraph{Java} We collected 100 projects from  the SF110 dataset~\citep{ICSE12}. We used EvoSuite~\citep{10.1145/2025113.2025179} to generate and run test cases and developed our tracing tool on top of Java Debugger~\citep{oracleJDB} to record the execution details of the projects. %We logged method invocations, variable values, and different program states during the execution cycle. 

\subsection{Data Filtering}
We collected whole program traces, from which we curated unique functions based on their entry and exit points in the execution trace logs. We excluded functions that only contain comments, too lengthy to fit into the models' context, and the functions which don't have meaningful functionality in their body. For example, some functions only contain one statement like  “return 0”, or “printf(“...”) or some functions are just a wrapper for another function which we have tested, such as "void myfunc()\{ func();\}". We obtained a total of 2125 Python, 876 C and 875 Java unique functions, with the sizes ranging from 3 to 516 lines of code. From these unique functions, we curated our task-specific datasets. 

In real-world code, we face many complexities, e.g., the input of a function and the values of a variable can be complex types. As the first step of probing models to reason about fine-grained code semantics for real-world code, we focus on ground truth values of primitive data types in all tasks, including {\tt int}, {\tt float}, {\tt str}, {\tt bool}, {\tt list}, {\tt pointer}, {\tt double}, {\tt dictionary}, {\tt tuple}, etc.  In the evaluation, we show that even for values of primitive types, the models find it challenging to predict them. We collected a total of 4483 samples from Python, C and Java, and constructed the ground truth for the above tasks and used for evaluation. See Table~\ref{tab:sample}.

\begin{table}[ht]
\centering
\caption{Number of Samples for Tasks}
\footnotesize
\label{tab:sample}
  \resizebox{0.55\textwidth}{!}{
\begin{tabular}{lrrrr}
\toprule
Task & Python & C & Java & Total Samples \\
\midrule
Task 1: Block & 1860 & 731 & \multicolumn{1}{c}{--} & 2591 \\
Task 1: Function & 308 & 94 & 74 & 476 \\
Task 2/Task 4: Statement  & 545 & 485 & \multicolumn{1}{c}{--} & 1030 \\

Task 3-1/Task 4: Loop & 105 & \multicolumn{1}{c}{--} & \multicolumn{1}{c}{--} & 105 \\
Task 3-2: Pointer & \multicolumn{1}{c}{--} & 49 & \multicolumn{1}{c}{--} & 49 \\
Task 3-3: Branch & 232 & \multicolumn{1}{c}{--} & \multicolumn{1}{c}{--} & 232 \\
\midrule
Total Samples & 3050 & 1359 & 74 & 4483 \\
\bottomrule
\end{tabular}
}
\end{table}

\section{Evaluation}
\label{Evaluation}
%\wei{Roy please add experimental setup and models} Selecting NL prompt, a list of models
%We probe five templates of prompts and used a validation dataset and selected the prompt for a particular model using validation set, we observe template are model sensitive, but not task sensitive. So we select a template for each model. In the following, we presented interesting experimental results. The complete results are available in our data package.

%\wei{Roy: you need to write a short summary and finding for each RQ}
%\wei{Results pending: we need gpt 4.o results, and also plot existing results for claude, gemini for all the figures}

%\TODO{1. Finish the evaluation setup}

%\TODO{2. replot all figures}

%\TODO{3. follow my template to write text for RQ2-RQ6}

%\subsection{Evaluation Setup}

%\fakeparagraph{Code LLMs} \TODO{make a table to introduce each model information and suggest a shrinked name or ID for each model} 

%Selecting NL prompt, a list of models
%We probe five templates of prompts and used a validation dataset and selected the prompt for a particular model using validation set, we observe template are model sensitive, but not task sensitive. So we select a template for each model. In the following, we presented interesting experimental results. The complete results are available in our data package.

We evaluate 14 SOTA LLMs, 8 reasoning models and 6 non-reasoning models (Table~\ref{tab:model-names} for full names and short IDs used in figures), including open-source models (Llama, phi), close-source/API models (GPT-4.0 Mini, Claude 3.5 and Gemini 1.5) and distilled models (DeepSeek R1 series), with the model parameter sizes ranging from 7 to 14 billions. We utilized vLLM(v0.3.3) as our inference engine to run the models.

\begin{table}[h]
\centering
\caption{Model Names and Their IDs in the figures }\label{tab:model-names}
\centering
\footnotesize
\setlength{\tabcolsep}{1pt}
  \resizebox{0.6\textwidth}{!}{
\begin{tabular}{@{}ll@{}}
\toprule
\textbf{Full Model Name} & \textbf{Model ID in the figures}\\
\midrule
openai/gpt-4.0-mini & GPT-4o (Reasoning) \\
anthropic.claude-3-5-sonnet-20241022-v2:0 & CL-3.5 (Reasoning) \\
gemini-1.5-flash-002 & Gem-1.5 (Reasoning) \\
meta-llama/Llama-3.1-8B-Instruct & L-3.1\\
Qwen/Qwen2.5-14B-Instruct-1M & Q-2.5\\
Qwen/Qwen2.5-Coder-7B-Instruct & Qwen2.5-C\\
deepseek-ai/DeepSeek-Coder-V2-Lite-Instruct & DS-C\\
microsoft/Phi-4-mini-instruct & Phi-4\\
microsoft/Phi-3.5-mini-instruct & Phi-3.5\\
ibm-granite/granite-3.2-8b-instruct & Gr-3.2 (Reasoning)\\
deepseek-ai/DeepSeek-R1-Distill-Qwen-7B & DSR1-Q-7B (Reasoning) \\
deepseek-ai/DeepSeek-R1-Distill-Llama-8B & DSR1-L (Reasoning)\\
deepseek-ai/DeepSeek-R1-Distill-Qwen-14B & DSR1-Q-14B  (Reasoning)\\
ibm-granite/granite-3.2-8b-instruct-preview & Granite-3.2 Pr (Reasoning)\\

\bottomrule
\end{tabular}
}
\end{table}

We designed five different natural language prompt templates (see Appendix/data package), and ran them on a sampled dataset for each model.  We observed that prompt templates are model-sensitive, but not task-sensitive. So we select a template for each model for all the tasks. We prompted the models to give a response inside specific tags (<ans> </ans>) and considered the response inside that tag to compare with the ground truth, as done in~\citep{gu2024cruxevalbenchmarkcodereasoning}. For our evaluation metrics, we used accuracy (exact matching of the generated outputs of the models and the ground truth label). %as done in 

%\wei{Roy add reference: as done in~\cite{XX}}. \roy{Cruxeval's <ans></ans> tag approach was same, but their evaluation metrics is different}

%We evaluated a subset of our benchmarks ~\ray{unfinished sentence}

In the following, we presented a selection of interesting results. 
For clarity, we present results from one representative LLM per model family to ensure model diversity. Please refer to the Appendix for the complete set of experimental results.

%selected the prompt for each model based on the results of a validation dataset. %We selected the best-performing natural language template from the validation dataset and used the template for all the tasks. \\  configuration see Appendix XX).

%\fakeparagraph{Evaluation Process and Metrics}
%\TODO{describe the evaluation details, such as prompts, LLM hyperparameters, and others}
\mycomment{
We designed five different natural language prompts (have to mention section of appendix/data package), and ran the prompts on a validation dataset prepared from the statement prediction dataset (RQ1). We selected the best-performing natural language template from the validation dataset and used the template for all the tasks. \\
We utilized vLLM(v0.3.3) as our inference engine, with the configuration parameters in Appendix XX. For our evaluation metrics, we used accuracy (exact matching of the generated output of the models and the ground truth label). We prompted the models to give a response inside specific tags (<ans> </ans>) and considered the response inside that tag to compare with the ground truth.}

\subsection{Results for RQ1: Block-level code semantics}
\mycomment{
\begin{itemize}
    \item Almost all the models are showing a gradual decrease in accuracy when the block size is increased. That indicates that models find it difficult to keep track of the run-time variable progression in the code.
    \item For C, the trend is missing because the sample size was too low.
    \item \figref{fig:rq2_2}, input predicition is very low, model doesn't get affected by the code sizes, they just can not infer model input by given output (which is a hard task). For output prediction, For some models we see the expected trend, of small codes having higher accuracy and larger models have lower, however, the plot is for Python, and in python sometimes, small codes can perform more things than larger codes (like comprehension makes the code smaller)
\end{itemize}
}

% In \figref{fig:rq1}, we plot input/output prediction results for code blocks and functions (a special case of code-block) of different sizes. In \figref{fig:rq1_block}, blue, orange, and green bars are for the results of predicting ouput of a block consisting of  one, two, and three statements, respectively.%\wei{Roy-presentation: the caption in Figure 6, small medium and large}

\begin{figure}[h]
    \centering
    \begin{subfigure}[b]{\textwidth} 
        \includegraphics[width=0.48\textwidth]{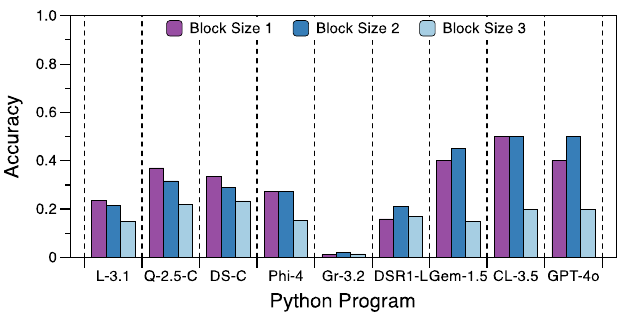}
        ~
        \includegraphics[width=0.48\textwidth]{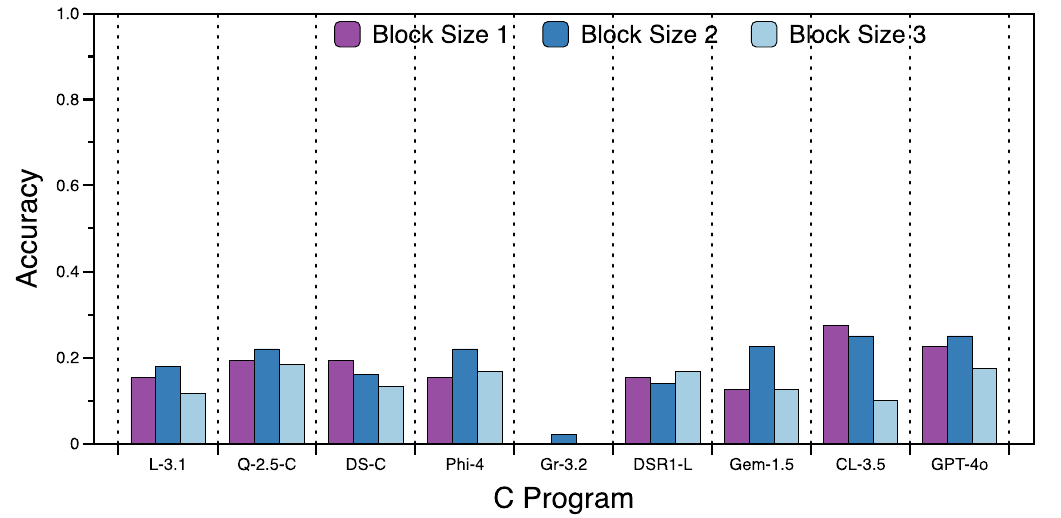}
        \caption{Output prediction for different {\em block} sizes}
        \label{fig:rq1_block}
    \end{subfigure}
    ~
    \begin{subfigure}[b]{\textwidth} 
        \includegraphics[width=0.48\textwidth]{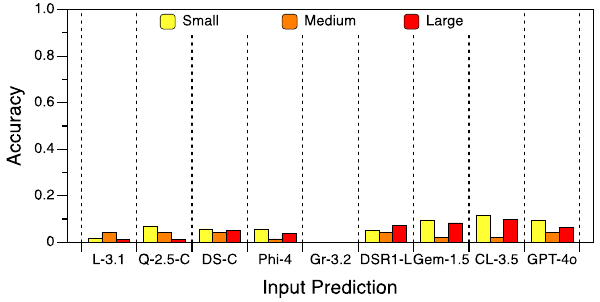}
        ~
        \includegraphics[width=0.48\textwidth]{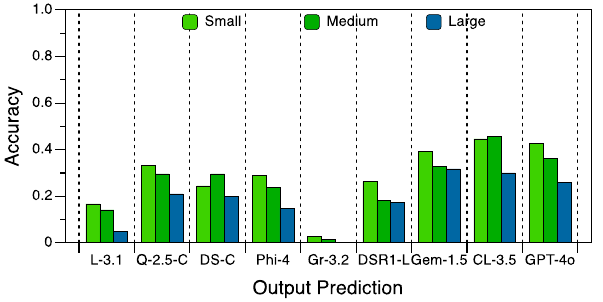}
        \caption{In/Out prediction for different {\em function} sizes}
        \label{fig:rq1_function}
    \end{subfigure}
    \caption{\textbf{RQ1:} Does increasing the size of code increase the difficulty of code reasoning?}
    \label{fig:rq1}
\end{figure}

\figref{fig:rq1} shows input/output prediction results for code blocks and functions (a special case of code block) across varying sizes. In \figref{fig:rq1_block} shows the results for three block sizes - blocks containing one, two, and three statements, respectively.

Overall, we observe that model accuracy is low even for small code blocks. For example, in C dataset, models such as Claude 3.5 and GPT-4o-mini achieve under 30\% accuracy on single-statement blocks. Python yields slightly better results, though no model exceeds 50\% accuracy. Performance further declines as block size increases from 1 to 3 statements, with open-source models performing significantly worse. This degradation stems from two primary challenges: models often fail to reason about individual statements, and they struggle to track variable state across statements. Notably, even Claude 3.5 achieves only 20\% accuracy on 3-statement Python blocks, and less than 10\% on C. However, in some cases, smaller blocks can be harder because they contain API calls.

A similar trend is observed in \figref{fig:rq1_function} (right: Output Prediction), where models perform better on smaller functions than larger ones for output prediction. However, performance on input prediction remains consistently poor (left figure) across all function sizes. This highlights a broader limitation: LLMs  are even less capable of understanding the "reverse" of operational semantics, i.e., inferring inputs from outputs. 
Even the best-performing model, Claude 3.5, achieves only around 12\% accuracy in input prediction for small functions.

%We observed a similar trend in \figref{fig:rq1_function} on the right, where predicting output for small functions is easier than the larger functions.  The models perform very badly for the input prediction tasks for all functions, independent of their sizes. The LLMs  are even less capable to understand the "reverse" of language operational semantics given an output of a statement, predicting its input. We see Claude 3.5 performs the best for input/output predictions among all the models.

\mycomment{
, and \figref{fig:rq2_2} representing input/output accuracy depending on code length. The left and right subplots in \figref{fig:rq2_1} denote Python and C accuracies, respectively, whereas in \figref{fig:rq2_2}, the left and right subplots show input and output predictions.\\
As the block size increases, most models generally tend to drop accuracy; however, even for block-1, models achieve lower accuracy than 40\%  in Python and 30\%  in C. For this task, we found that reasoning models are not performing as well as the non-reasoning models. Overall, when comparing two languages, Python consistently shows comparatively better results than C.
We observed that some models perform poorly on the output prediction task as the code length increases. For output prediction, the cluade outputperform other models for small and medium-sized codes, achieving more than 40\% accuracy. On the other hand, irrespective of code length, input prediction is hard for the models and having less than 20\% accuracy across all the models.
}

% \begin{figure}[htbp]
%     \centering
%     \begin{subfigure}[b]{0.48\textwidth} 
%         \includegraphics[width=\textwidth]{Fig/rq2_python.pdf}
%      %   \caption{Python}
%     \end{subfigure}
%     \hfill
%     \begin{subfigure}[b]{0.48\textwidth}
%         \includegraphics[width=\textwidth]{Fig/rq2_c.pdf}
    
%     \end{subfigure}
%     \caption{\textbf{RQ1 (block):} Does increasing the size of code increase the difficulty of code reasoning?}
%     \label{fig:rq2_1}
% \end{figure}

% \begin{figure}[htbp]
%     \centering
%     \begin{subfigure}[b]{0.48\textwidth} 
%         \includegraphics[width=\textwidth]{Fig/rq2_input.pdf}
%      %   \caption{Python}
%     \end{subfigure}
%     \hfill
%     \begin{subfigure}[b]{0.48\textwidth}
%         \includegraphics[width=\textwidth]{Fig/rq2_output.pdf}
%       %  \caption{C}
%     \end{subfigure}
%     \caption{\textbf{RQ1 (function):} It is not clear here how function and blocks are different}}
%     \label{fig:rq2_2}
% \end{figure}

% \begin{figure}[htbp]
%     \centering
%     \includegraphics[width=0.95\textwidth]{RQ2_Results/output_accuracy_no_length.png}
%     \caption{\textbf{RQ2:} input/output prediction accuracy changes based on code length }
%     \label{fig:rq2_image3}
% \end{figure}

% \begin{figure}[htbp]
%     \centering
%     \includegraphics[width=0.95\textwidth]{RQ2_Results/input_accuracy_no_length.png}
%     \caption{\textbf{RQ2} Input Prediction Accuracy based on code length}
%     \label{fig:rq2_image3}
% \end{figure}

\subsection{Results for RQ2: Statement-level code semantics}
%\wei{Ray: we need a box plot here like Cruxeval paper?}
\mycomment{
\begin{itemize}
    \item For the models, it's hard for them to predict the execution of function calls, we sampled third party modules(that can be installed by pip), which are common modules like os, sys, time, math etc,) and yet the accuracy on Python is low. 
    \item Some models failing to predict about boolean execution is also surprising, as we are giving the expression variable values explicitly.
\end{itemize}
}

\begin{figure}[htbp]
    \centering
    \begin{subfigure}[b]{0.48\textwidth}
        \includegraphics[width=\textwidth]{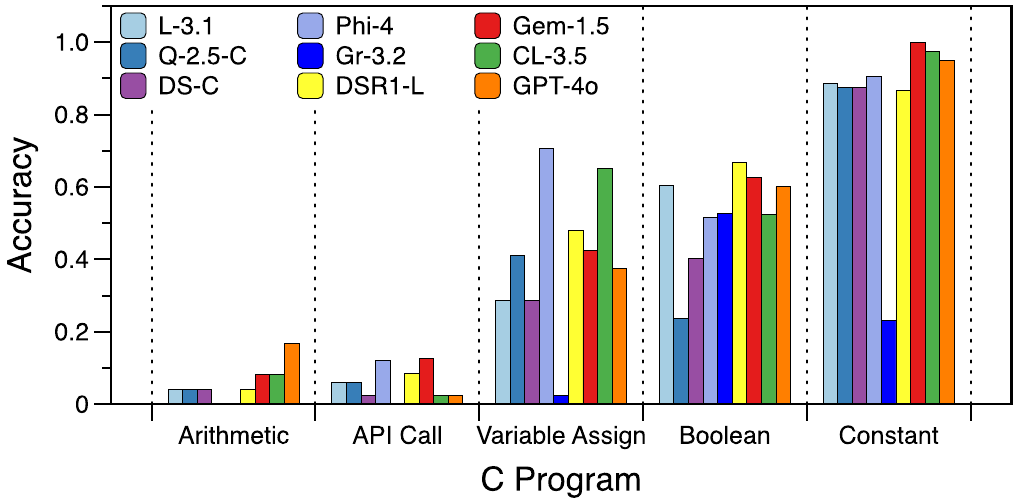}
      %  \caption{C}
        \label{fig:subfig2}
    \end{subfigure}
    \hfill
    \begin{subfigure}[b]{0.48\textwidth} 
        \includegraphics[width=\textwidth]{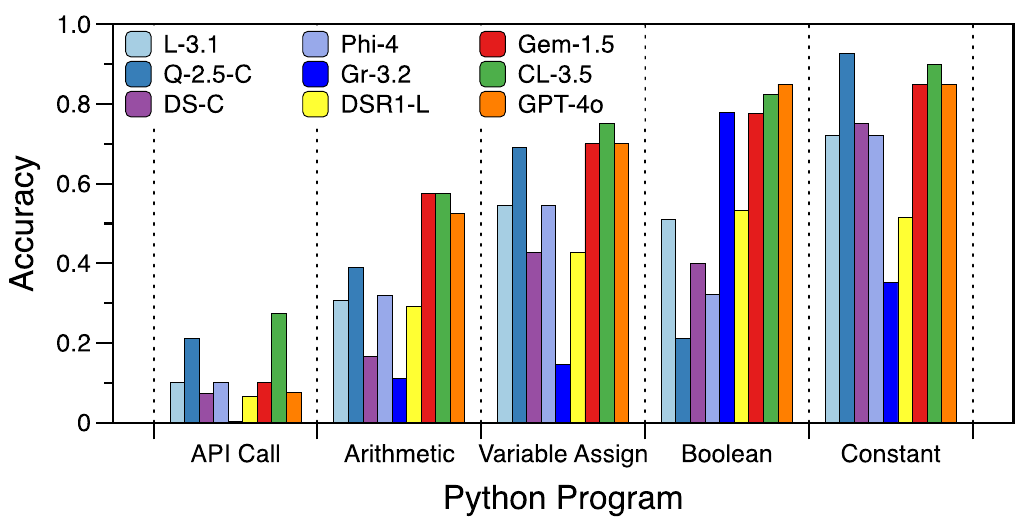}
     %   \caption{Python}
        \label{fig:subfig1}
    \end{subfigure}
    \caption{\textbf{RQ2:} What types of program statements can the model understand well?}
    \label{fig:rq1_side_by_side}
\end{figure}
%\wei{Roy: yellow bar in this figure can move to the left, it seems quite high}
%\ray{Can we show an overall result across all statement}

%\TODO{1. describe the figure information}
%\ray{We need to say what exactly we have done, what does statement level prediction mean?}

As shown in RQ1, models struggle with value prediction even for single statements (e.g., block size 1). In RQ2, we further analyze model performance by categorizing results based on statement types. \figref{fig:rq1_side_by_side} presents these results, with the left plot showing C and the right showing Python. Each plot groups model performance by statement type to highlight specific areas of strength and weakness.

% RQ1 we have seen that models suffer from value prediction for even single statement (see output prediction for block size 1).
% In RQ2, we deep-dive to evaluate the model's performance; we group the results of each model according to the type of statement.
% In \figref{fig:rq1_side_by_side}, we show results for RQ2, left for the Python dataset and right for C. 
% In each plot, we group the results of each model according to the type of statement. 
% including \texttt{Function Call}, \texttt{Arithmetic}, \texttt{Variable}, \texttt{Boolean}, and \texttt{Constant}.

%\TODO{2. describe the findings}
%We observe that models still face significant challenges to reason about single  statement like 
We observe that arithmetic and API/calls are the most statement types, even for the best reasoning models like GPT4.0-mini and Claude 3. For APIs, we sampled frequently used third-party libraries, like {\tt os}, {\tt sys}, and {\tt math} installed by {\tt pip}, but the models do not have knowledge about their execution semantics. We also experimented with adding the API definitions in the prompt, but it didn't increase the performance significantly \ref{app:api}. Models handle better for predicting Boolean values such as the output of a comparison statement, and also statements where constant is assigned to a variable. The models may understand the assignment operator "=" and have captured easy patterns like "a=3 indicates {\tt a} has value 3 after executing this statement". We did not observe significant advantage of reasoning models over non-reasoning models  for this task.

\subsection{Results for RQ3: Code properties within a function}

\mycomment{
\begin{itemize}
    \item Despite of being reasoning models, granite doesn't work in 0-shot settings. 
    \item Reasoning and non-reasoning models excel while they need to predict categorise or "yes"/"no" binary prediction.
\end{itemize}
}

\begin{figure}[htbp]
    \centering
    \begin{subfigure}[b]{0.32\textwidth} 
        \includegraphics[width=\textwidth]{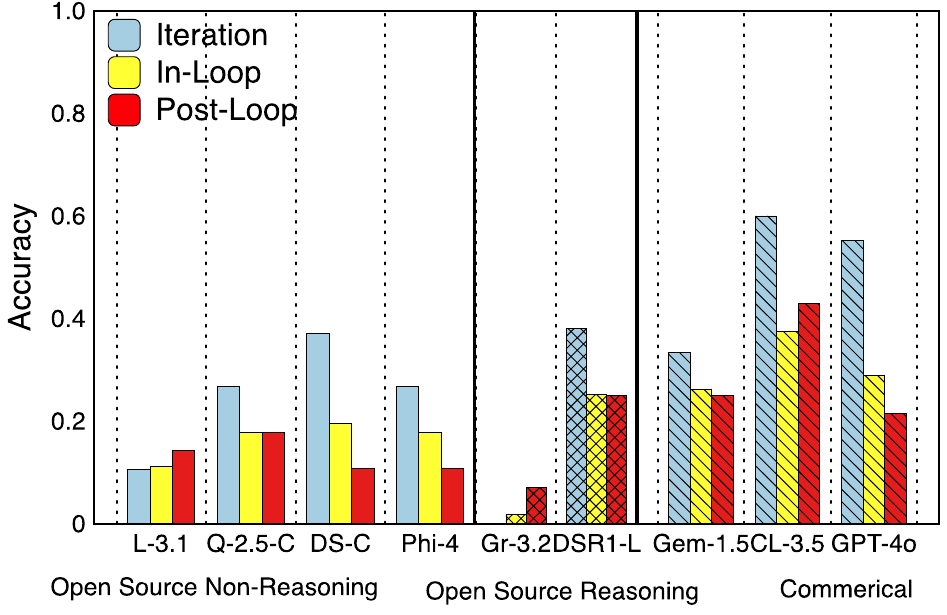}
    \caption{Loop Properties}
    \label{fig:rq3_loop}
    \end{subfigure}
    ~
    \begin{subfigure}[b]{0.32\textwidth}
        \includegraphics[width=\textwidth]{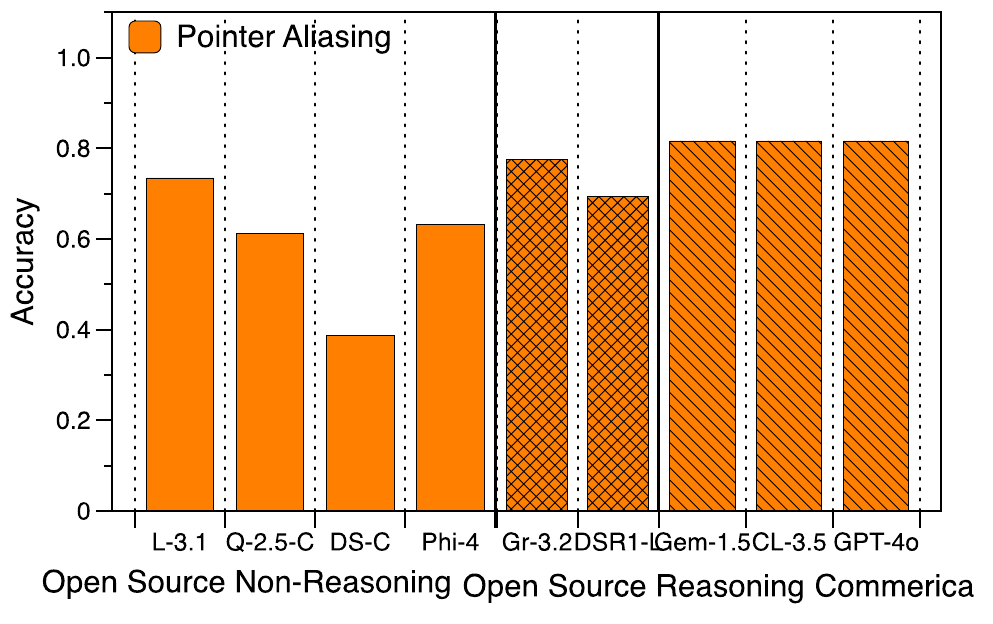}
        \caption{Pointer Aliasing}
        \label{fig:rq3_alias}
    \end{subfigure}
    ~
    \begin{subfigure}[b]{0.32\textwidth}
        \includegraphics[width=\textwidth]{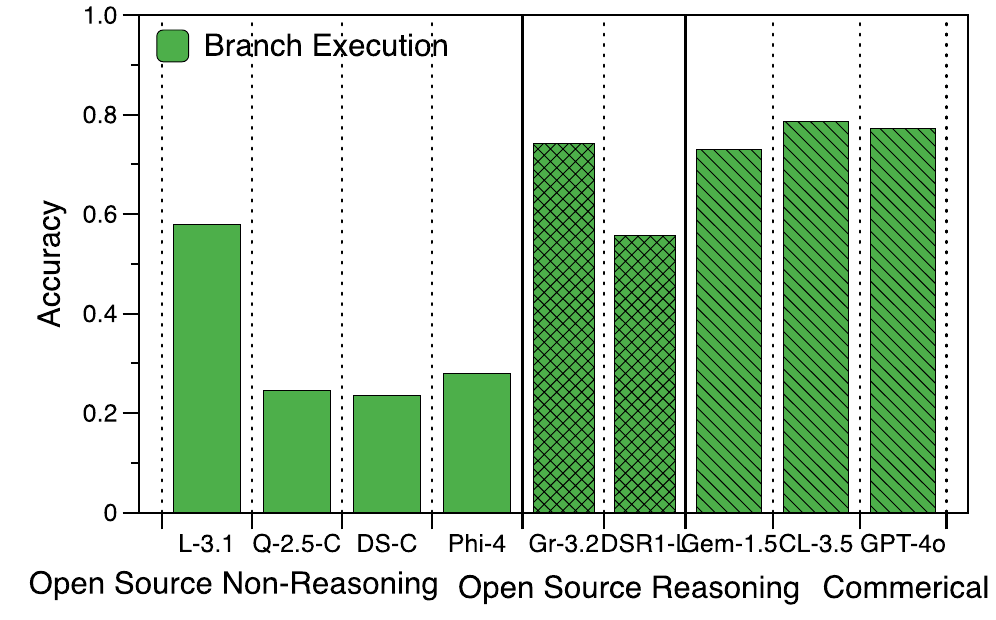}
        \caption{Branch Execution}
        \label{fig:rq3_branch}
    \end{subfigure}
    \caption{\textbf{RQ3:} Can models reason about different program properties?}
    \label{fig:rq3}
\end{figure}

In \figref{fig:rq3_loop}, we report results on predicting the number of loop iterations, values in and after the loop, given the input of the function. We observe that models feel difficult to predict values after executing the loop. Loop iterations are the easiest tasks among the three. Our intuition is that sometimes certain patterns in the code text are linked to the loop iterations. For example, Python code {\tt for i in range(100):}. implies that loop iteration is 100. Somehow, some models know these patterns and constants are linked to the loop iterations. We inspected the predicted loop values and did not find a trend that the models just use any constant numbers in the code text as their answers. 

%\wei{Roy: when wrong, does the in and loop values often predict constant values appeared in the loop?}

In \figref{fig:rq3_alias} and \figref{fig:rq3_branch}, we show that given an input, predicting pointer aliasing at a program location and whether a branch can be taken is easier than loop properties. Here, the models only need to predict "yes"/"no". The models predict pointer aliasing better than branch execution. We believe that the models are able to connect code patterns such as "p=q" to the aliasing definition provided in our prompt "when two pointers store the same memory addresses, they are aliasing". 
Notably, some open-source models perform below 50\% on these binary classification tasks—worse than random guessing.

\mycomment{
In the case of loop property prediction, the majority of the models show comparatively better performance in loop iteration prediction. The highest loop iteration accuracy is achieved by Claude-3.5-sonnet, which is about 60\%. However, models exhibit lower performance in the case of in-loop and post-loop value prediction, with the post-loop value being the most challenging task. For in-loop prediction, Claude-3.5-sonnet achieves the highest performance 40\% and, 45\% for in-loop and post-loop predictions, respectively. The granite models showed minimal capability for reasoning about loop properties among all the models.

For aliasing and branch prediction, we prompted the models in binary prediction (yes/no) settings, and found out the models handle simpler binary classification tasks effectively. For aliasing prediction, nearly all the models except deepseek coder v2 achieve accuracy over 60\%. Among them, deepseek-R1-Qwen-14 outperforms all models by reaching an accuracy of 84\%. Similarly, in branch prediction, the majority of models exceed 60\% accuracy, with Qwen 2.5-14B achieving the highest accuracy of 80\%.
}

% \begin{figure}[htbp]
%     \centering
% \includegraphics[width=0.66\textwidth]{Fig/RQ3-Loop.pdf}
%     \caption{\textbf{RQ3 (function):} Can models reason about loop properties?}
%     \label{fig:rq3_image1}
% \end{figure}

% \begin{figure}[htbp]
%     \centering
%     \begin{subfigure}[b]{0.48\textwidth} 
%         \includegraphics[width=\textwidth]{Fig/RQ3-Alias.pdf}
%      %   \caption{Python}
%     \end{subfigure}
%     \hfill
%     \begin{subfigure}[b]{0.48\textwidth}
%         \includegraphics[width=\textwidth]{Fig/RQ3-Branch.pdf}
%       %  \caption{C}
%     \end{subfigure}
%     \caption{\textbf{RQ3:} Can models predict pointer aliasing and branch execution?}
%     \label{fig:rq3_image2}
% \end{figure}

% \begin{figure}[htbp]
%     \centering
%     \includegraphics[width=0.95\textwidth]{RQ3_Results/aliasing_accuracy.png}
%     \caption{\textbf{RQ3 (function):} How do models perform on reasoning about pointer properties in C code?}
%     \label{fig:rq3_image2}
% \end{figure}

% \begin{figure}[htbp]
%     \centering
%     \includegraphics[width=0.95\textwidth]{RQ3_Results/condition_accuracy.png}
%     \caption{\textbf{RQ3 (function):} How do models perform on branch prediction task?}
%     \label{fig:rq3_image2}
% \end{figure}

\subsection{Results for RQ4: Different prompting techniques}

\mycomment{
\begin{itemize}
    \item Providing same function as in-context example helps, but CoT doesn't, For statement prediction, maybe the model doesn't need step by step approaches for the statement prediction task, because it doesn't to go through the whole code, just need to reason about the statement (CoT working as redundant?)
\end{itemize}
}

\begin{figure}[htbp]
    \centering
    \begin{subfigure}[b]{\textwidth} 
        \includegraphics[width=0.48\textwidth]{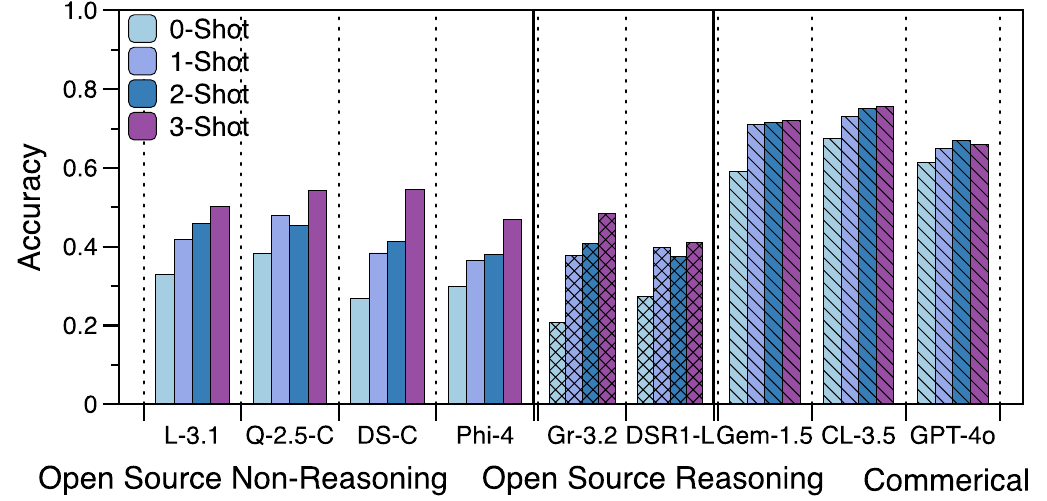}
        \includegraphics[width=0.48\textwidth]{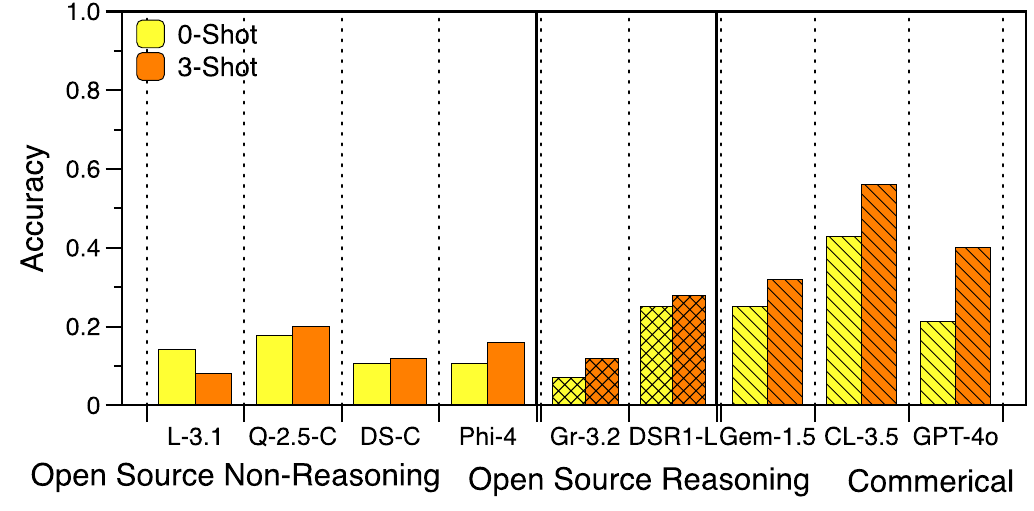}
        \caption{In-context Learning: (left) statement prediction, (right) post-loop prediction}
        \label{fig:rq4_incontext}
    \end{subfigure}
    ~
    \begin{subfigure}[b]{\textwidth} 
        \includegraphics[width=0.32\textwidth]{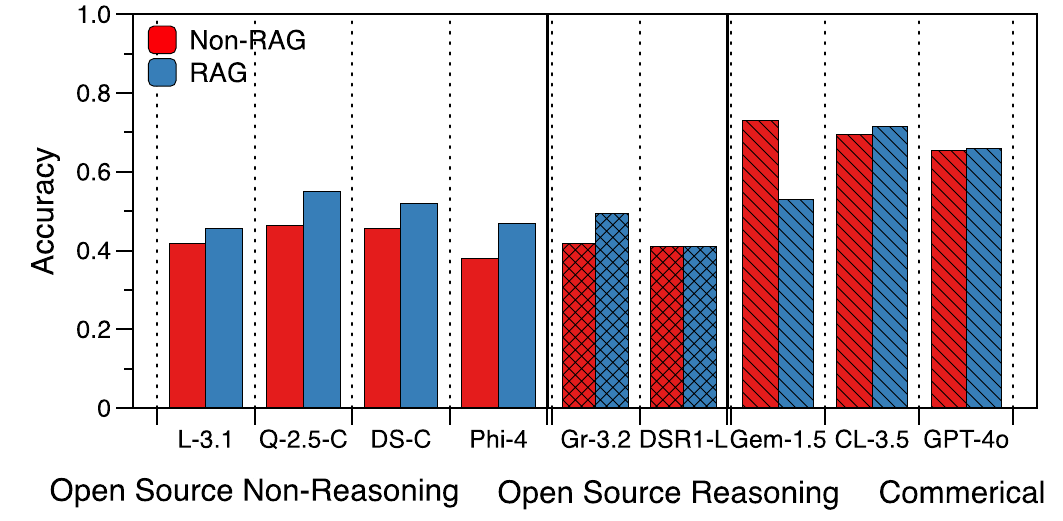}
        \includegraphics[width=0.32\textwidth]{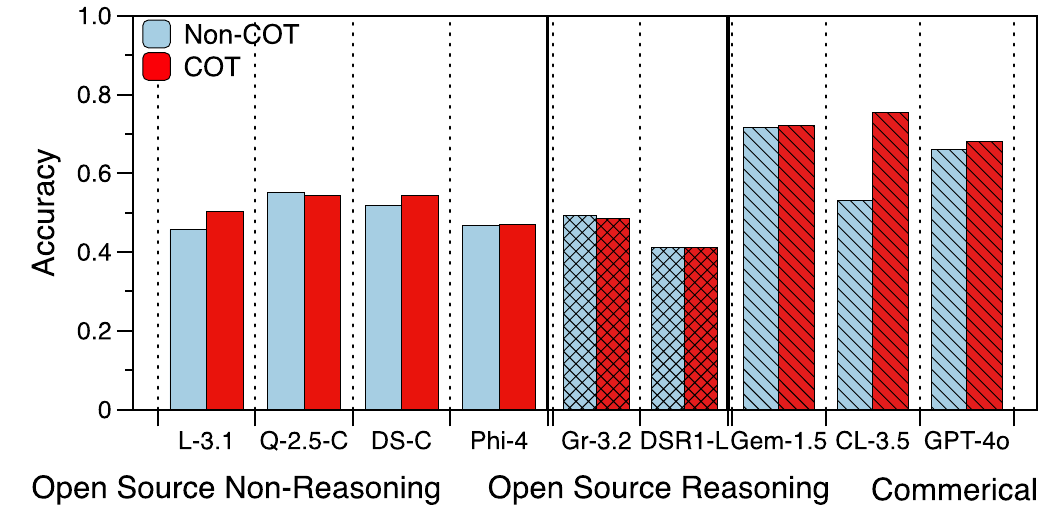}
        \includegraphics[width=0.32\textwidth]{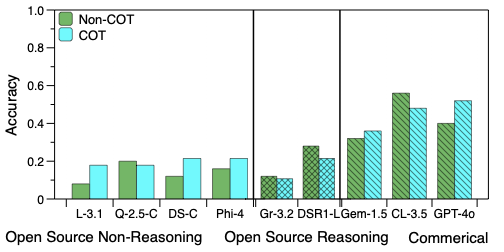}
        \caption{RAG \& COT: (left) comparing random shots and shots relevant to the query for statement prediction, \\
        (mid) COT for statement prediction, (right) COT for post-loop prediction}
        \label{fig:rq4_cot}
    \end{subfigure}
    \caption{\textbf{RQ4:} Can different prompting strategies help?} %~\ray{show in figure which one is statement and which one is loop. What is different and same?}}
    \label{fig:rq4}
\end{figure}

In \figref{fig:rq4}, we show results of different prompting techniques on statement prediction and loop property predictions (relatively difficult tasks in our list). Our results show that in both cases, models benefited from more shots in the prompt. When we prompt models and provide examples more relevant to the query (RAG style); that is, for statement prediction, we provide shots with the same type of statement, and for loops, we provide shots of different loops in the same function, models improved their performance. However, applying a simple COT by "asking models to think step by step" at the beginning of the prompt did not help much for statement prediction, but helped for loop property prediction for some models. Our intuition is that compared to statement prediction, loop reasoning, e.g., predicting values after a loop, may be more complex and can benefit from  multi-step reasoning.
%\wei{we don't have loop figure}

%\wei{Figure 10: add COT on post-loop (figure 9, right) individual statement may need less COT, but the loop needs multi-stage reasoning more}

\mycomment{
shows the result of adding different numbers of in-context examples in the RQ1 statement prediction dataset. It is evident from the figure that adding in-context examples consistently enhances model performance. This suggests that models benefited from adding contextual information in the prompt. Next, we tried to investigate how models perform if we select the in-context examples in a more controlled setting. For that, we selected in-context examples belonging to the same function as the query function, but we ensured that they involved different types of statement prediction. \figref{fig:rq4_image2} shows the result of this experiment, and we can see that adding the same in-context examples help the model to reason more effectively, enhancing the overall performance. After that, we also experimented with CoT prompting, and the result is shown in \figref{fig:rq4_image2}. We found out that adding CoT didn't yield better performance in general for all the models.
}

% \begin{figure}[htbp]
%     \centering
%     \begin{subfigure}[b]{0.48\textwidth} 
%         \includegraphics[width=\textwidth]{Fig/RQ4-1-Left.pdf}
%         % \caption{Python}
%     \end{subfigure}
%     \hfill
%     \begin{subfigure}[b]{0.48\textwidth}
%         \includegraphics[width=\textwidth]{Fig/RQ4-1-Right.pdf}
%     \end{subfigure}
%     \caption{\textbf{RQ4 (prompt)}: Can RAG and in-context learning help?}
%     \label{fig:rq4_incontext}
% \end{figure}

% %\wei{Roy: if we have time, add three shots for loop properties}

% \begin{figure}[htbp]
%     \centering
%     \begin{subfigure}[b]{0.48\textwidth} 
%         \includegraphics[width=\textwidth]{Fig/RQ4-2-Left.pdf}
%         % \caption{Python}
%     \end{subfigure}
%     \hfill
%     \begin{subfigure}[b]{0.48\textwidth}
%         \includegraphics[width=\textwidth]{Fig/RQ4-2-Right.pdf}
%     \end{subfigure}
%     \caption{\textbf{RQ4 (prompt)}: Can RAG and COT help?}
%     \label{fig:rq4_cot}
% \end{figure}

% \begin{figure}[htbp]
%     \centering
%     \includegraphics[width=0.95\textwidth]{RQ4_Results/Concrete_value_incontext_comparison}
%     \caption{{\bf RQ4 (prompt) }: Does using similar examples/queries as shots help?}
%     \label{fig:rq4_image2}
% \end{figure}

% \begin{figure}[htbp]
%     \centering
%     \includegraphics[width=0.95\textwidth]{RQ4_Results/Concrete_COT_comparison.png}
%     \caption{{\bf RQ4 (prompt) }: Does simple COT prompt help?}
%     \label{fig:rq4_image3}
% \end{figure}

\subsection{Results for RQ5: Approximation of code semantics}
\begin{figure}[htbp]
    \centering
    \begin{subfigure}[b]{0.32\textwidth} 
        \includegraphics[width=\textwidth]{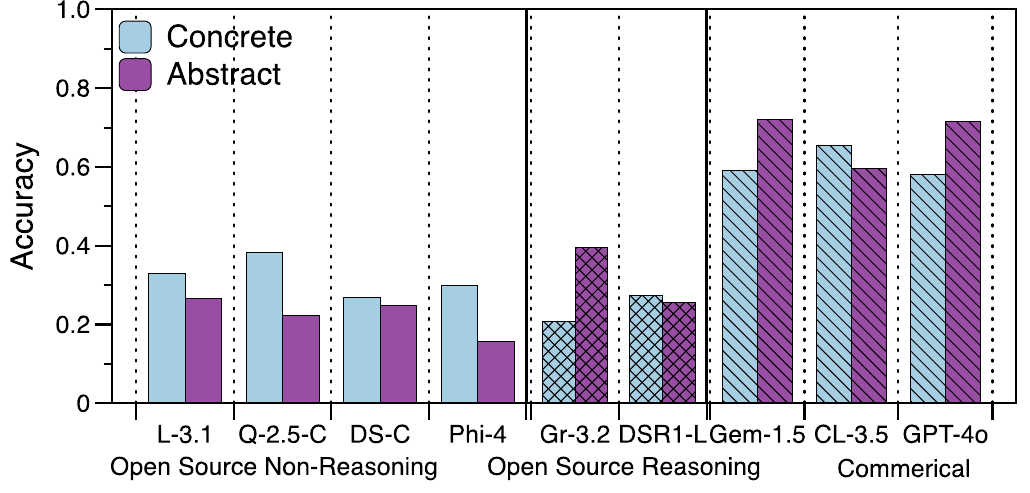}
       \caption{0-shot statement prediction} \label{fig:rq6_image11}
    \end{subfigure}
    \hfill
    \begin{subfigure}[b]{0.32\textwidth}
        \includegraphics[width=\textwidth]{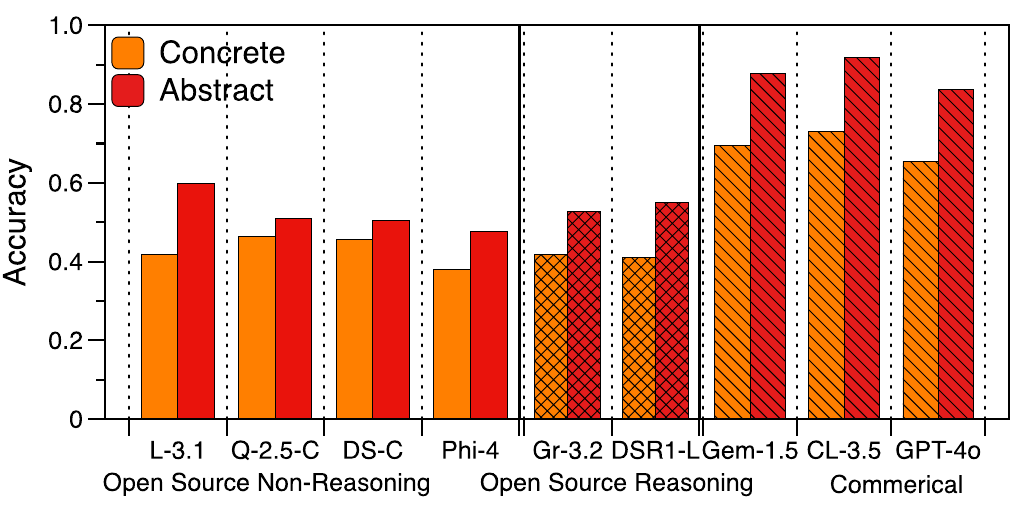}
        \caption{3-shot statement prediction} \label{fig:3s}
    \end{subfigure}
        \begin{subfigure}[b]{0.32\textwidth}
        \includegraphics[width=\textwidth]{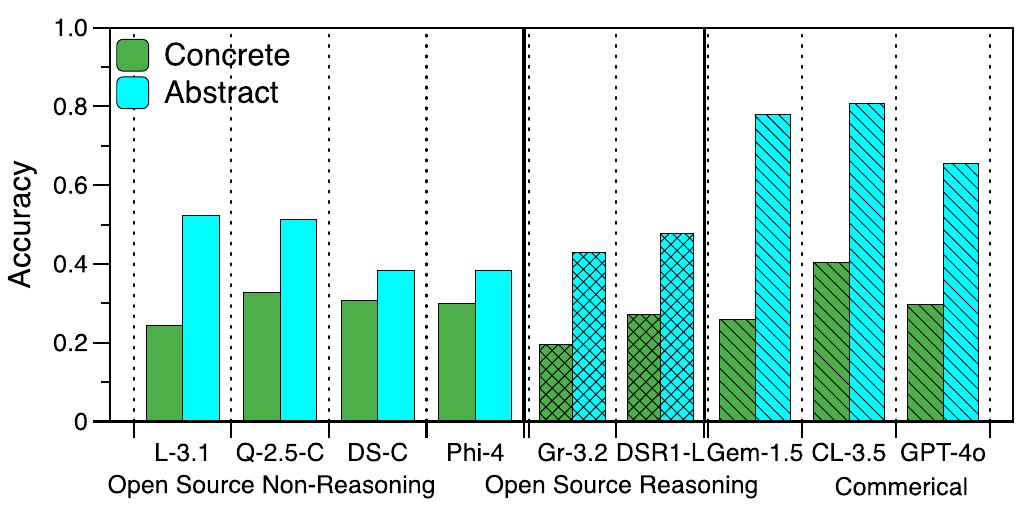}
        \caption{3-shot post-loop prediction}
        \label{fig:3l}
    \end{subfigure}
    \caption{\textbf{RQ5:} Can models reason about an approximation of code semantics? (Python results)}
    \label{fig:rq6_image1}
\end{figure}

In \figref{fig:rq6_image1}, we observed that the models reported better performance to predict an approximation of code semantics, for both statement (\figref{fig:3s}) and loop (\figref{fig:3l}) predictions. See also Table~\ref{tab:random baseline}  in Appendix~\ref {app:RQ5 results} for comparing with random baselines. Interestingly, when we provide only the definition of "abstract" values (mapping from a range of concrete values to an abstract value) in the prompt , without giving an example showing an "abstract" output for a given input, the models cannot predict abstract values better than concrete values (\figref{fig:rq6_image11}). Most models failed to apply definitions directly to the query examples; however, when we provide 3-shots of examples in the prompt, all the models can predict abstract values better than concrete ones. 

\mycomment{
Models seem to perform better if we add in-context examples in the case of Quantized Prediction. Without in-context examples, models failed to predict the value in abstract format, though we provided the concrete value to abstract value mapping in the 0-shot prompt. When we are adding in-context examples with proper abstract response, the models are performing better than concrete value prediction. 
}

\subsection{Results for RQ6: Different programming languages}

\begin{figure}[htbp]
    \centering
    \begin{subfigure}[b]{0.48\textwidth} 
  \includegraphics[width=\textwidth]{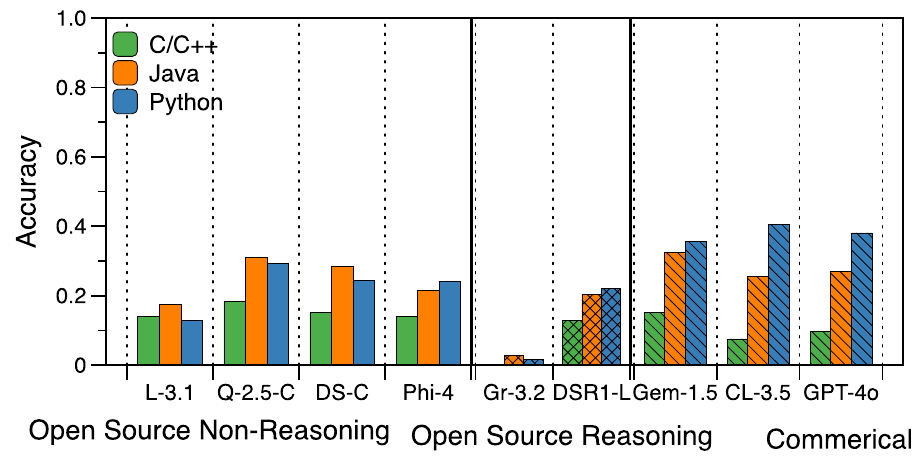}
       \caption{Output Prediction}
       \label{fig:out}
    \end{subfigure}
    \hfill
    \begin{subfigure}[b]{0.48\textwidth}
            \includegraphics[width=\textwidth]{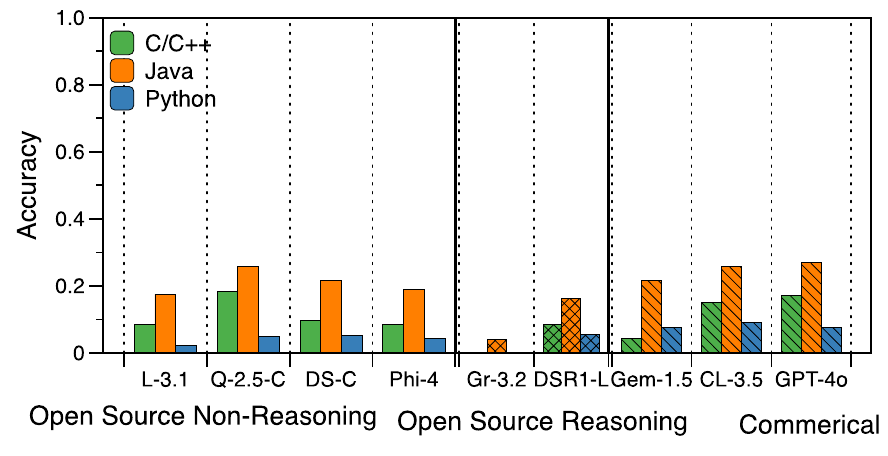}
      
       \caption{Input Prediction}
       \label{fig:in}
    \end{subfigure}
    \caption{\textbf{RQ6:} Any particular programming languages are easier for the models?}
    \label{fig:rq6}
\end{figure}

\mycomment{
\begin{itemize}
    \item Ensuring an unbiased dataset across all the languages was tough. The input prediction accuracy lower for python might be a reason for that.
\end{itemize}
}
Using input/output prediction as a case study, we investigated models code reasoning capabilities for different programming languages. \figref{fig:out} shows that Java and Python performed better than C when predicting output given input. Our intuition is that compared to the C code, Java and Python code are more high-level and closer to the natural languages than C; also {probably models have seen less C code than Python/Java code in the training data}. However, % \figref{fig:rq5_image2} shows that
the models reported the lowest accuracy for input prediction of Python code (\figref{fig:in}). %~\ray{what is the intuition here?}. 

\mycomment{
We present our RQ5 results in \figref{fig:rq5_image1} and \figref{fig:rq5_image2}, showing the input-output accuracy of the models in three different languages. For output prediction, Java exhibits overall higher accuracy than the other two languages for the majority of the models. Interestingly, Claude 3.5 sonnet gives the highest accuracy on python output prediction, which is 40\%. In contrast, the granite models showed the lowest output performance across all models.  Now, coming to the input prediction results, in line with the output prediction, Java shows superior performance than the other two languages. Surprisingly, Python ranked lowest in this task, and Granaite models continued to show the lowest performance of all the models.  
}

%\wei{ROy, Need to sort the output accuracy bars}

% \begin{figure}[htbp]
%     \centering
%     \includegraphics[width=0.95\textwidth]{RQ5_Results/output_prediction_language_comparison.png}
%     \caption{{\bf RQ5 (language):} Any particular programming languages are easier for the models?}
%     \label{fig:rq5_image1}
% \end{figure}

% \begin{figure}[htbp]
%     \centering
%     \includegraphics[width=0.95\textwidth]{RQ5_Results/input_prediction_language_comparison.png}
%     \caption{\textbf{RQ5} Input Prediction}
%     \label{fig:rq5_image2}
% \end{figure}

%\input{Table/related}

\section{Related Work}
%\wei{I have done a version, but need proofread}
\fakeparagraph{Code Reasoning Benchmarks}
CruxEval~\citep{gu2024cruxevalbenchmarkcodereasoning} assesses LLMs' performance on synthetic Python programs using the task of input-output prediction for a function. CruxEval-X~\citep{xu2024cruxeval} extends this work to multilingual settings by translating synthetic Python programs in CruxEval to other languages using LLMs. REval~\citep{chen2024reasoningruntimebehaviorprogram} evaluated branch prediction tasks using ClassEval and HumanEval~\citep{chen2021evaluatinglargelanguagemodels}. 
CodeMind~\citep{liu2024codemindframeworkchallengelarge} proposed output prediction and code synthesis tasks  on existing code benchmarks~\citep{chen2021evaluating, austin2021program, gu2024cruxevalbenchmarkcodereasoning, puri2021codenet}. 
% \wei{Roy, we need to add citations for these benchmarks Mbpp, humaneval, cruxeval, codenet}. 
They found that LLM reasoning capabilities deteriorate as program complexity increases~\citep{liu2024codemindframeworkchallengelarge, Zhang_2024}. 
Our benchmark \bench is the first that used real-world Python, C and Java code to evaluate LLMs' reasoning capabilities. While most code reasoning benchmarks reason about function level execution semantics, we proposed and made ground truth available for a spectrum of fine-grained reasoning tasks regarding program behaviors within a function.

\fakeparagraph{Other Code Application Benchmarks}
SWE-Bench~\citep{jimenez2024swebenchlanguagemodelsresolve} used the task of generating patches to resolve a given GitHub issues for real-world Python projects. SWE-PolyBench~\citep{rashid2025swepolybenchmultilanguagebenchmarkrepository} extends this work to other programming languages. KGym~\citep{mathai2024kgym} delivered a benchmark consisting of Linux kernel crash data and evaluated LLMs' capabilities of resolving Linux kernel crashes. These benchmarks focus on task-specific performance rather than fine-grained code semantics understanding. There are also benchmarks for code generation~\citep{li2024evocodebenchevolvingcodegeneration, zhang2024codeagentenhancingcodegeneration, 10.1145/3597503.3623316, chen2025dynamic, du2023classeval} and code completion~\citep{10.1145/3597503.3639138, ding2023crosscodeevaldiversemultilingualbenchmark}. However, most of these datasets—such as BigCodeBench~\citep{zhuo2024bigcodebenchbenchmarkingcodegeneration} and CodeBenchGen~\citep{xie2024codebenchgencreatingscalableexecutionbased} are restricted to a single language (primarily Python) and extracted from isolated competitive programming problems.

\section{Conclusions}
%\wei{Ray please proofread this}
Code semantic reasoning is foundational for solving many software engineering applications. We propose a novel code benchmark and dataset, CodeSense, extracted from 744 Python, C and Java real-world projects, for evaluating LLMs capabilities of code semantic reasoning. We defined a spectrum of fine-grained code reasoning tasks include value predictions at various granularities of the code and program properties prediction for important code constructs like loops, pointers and branches. We developed a framework and tools that can build, test and trace software projects in different programming languages, and can automatically generate ground truth for fine-grained code semantic reasoning tasks. We conducted a comprehensive study on SOTA LLMs. We found that models in general lack the knowledge of code semantics and face challenges for reasoning about even single statements. In limited cases, models can establish the correlation of code semantics description in natural language with some simple frequent code patterns. We hope our dataset and framework can enable further code semantic benchmarks and provide ground truth for future LLMs post-training.

\newpage
%evaluation and improvement for LLMs 

%Producing ground-truth for fine-grained code semantics tasks for real-world code is challenging. 

%In this paper, we introduce a real-world, fine-grained benchmark for evaluating the code semantic understanding capabilities of large language models (LLMs). Our dataset is constructed from real-world software projects, with ground truth labels generated through program execution to ensure semantic correctness. To enable detailed analysis, we define a suite of fine-grained tasks that capture diverse and nuanced aspects of code semantics. We evaluate a broad range of state-of-the-art code LLMs, including both open-source and commercial models, and our results show that these models still struggle to fully grasp the fine-grained semantics present in real-world code.
%This work makes a valuable contribution by identifying and formalizing the challenge of semantic understanding in practical software contexts, providing a high-quality dataset and benchmark to support systematic evaluation, and offering a comprehensive empirical study that highlights the current limitations of existing models. We hope this benchmark serves as a foundation for developing more semantically-aware and reliable code intelligence systems.

%\TODO{Evaluating fine-grained code semantics is important. Producing ground-truth for fine-grained code semantics tasks for real-world code is challenging. This paper makes available ...}

\newpage
\bibliographystyle{unsrtnat}
\bibliography{bibliography}

\begin{thebibliography}{40}
\providecommand{\natexlab}[1]{#1}
\providecommand{\url}[1]{\texttt{#1}}
\expandafter\ifx\csname urlstyle\endcsname\relax
  \providecommand{\doi}[1]{doi: #1}\else
  \providecommand{\doi}{doi: \begingroup \urlstyle{rm}\Url}\fi

\bibitem[Liu et~al.(2023)Liu, Xia, Wang, and Zhang]{liu2023codegeneratedchatgptreally}
Jiawei Liu, Chunqiu~Steven Xia, Yuyao Wang, and Lingming Zhang.
\newblock Is your code generated by chatgpt really correct? rigorous evaluation of large language models for code generation, 2023.
\newblock URL \url{https://arxiv.org/abs/2305.01210}.

\bibitem[Jain et~al.(2024)Jain, Han, Gu, Li, Yan, Zhang, Wang, Solar-Lezama, Sen, and Stoica]{jain2024livecodebenchholisticcontaminationfree}
Naman Jain, King Han, Alex Gu, Wen-Ding Li, Fanjia Yan, Tianjun Zhang, Sida Wang, Armando Solar-Lezama, Koushik Sen, and Ion Stoica.
\newblock Livecodebench: Holistic and contamination free evaluation of large language models for code, 2024.
\newblock URL \url{https://arxiv.org/abs/2403.07974}.

\bibitem[Zhuo et~al.(2024)Zhuo, Vu, Chim, Hu, Yu, Widyasari, Yusuf, Zhan, He, Paul, Brunner, Gong, Hoang, Zebaze, Hong, Li, Kaddour, Xu, Zhang, Yadav, Jain, Gu, Cheng, Liu, Liu, Wang, Lo, Hui, Muennighoff, Fried, Du, de~Vries, and Werra]{zhuo2024bigcodebenchbenchmarkingcodegeneration}
Terry~Yue Zhuo, Minh~Chien Vu, Jenny Chim, Han Hu, Wenhao Yu, Ratnadira Widyasari, Imam Nur~Bani Yusuf, Haolan Zhan, Junda He, Indraneil Paul, Simon Brunner, Chen Gong, Thong Hoang, Armel~Randy Zebaze, Xiaoheng Hong, Wen-Ding Li, Jean Kaddour, Ming Xu, Zhihan Zhang, Prateek Yadav, Naman Jain, Alex Gu, Zhoujun Cheng, Jiawei Liu, Qian Liu, Zijian Wang, David Lo, Binyuan Hui, Niklas Muennighoff, Daniel Fried, Xiaoning Du, Harm de~Vries, and Leandro~Von Werra.
\newblock Bigcodebench: Benchmarking code generation with diverse function calls and complex instructions, 2024.
\newblock URL \url{https://arxiv.org/abs/2406.15877}.

\bibitem[Xie et~al.(2024)Xie, Xie, Sheth, Liu, Fried, and Rose]{xie2024codebenchgencreatingscalableexecutionbased}
Yiqing Xie, Alex Xie, Divyanshu Sheth, Pengfei Liu, Daniel Fried, and Carolyn Rose.
\newblock Codebenchgen: Creating scalable execution-based code generation benchmarks, 2024.
\newblock URL \url{https://arxiv.org/abs/2404.00566}.

\bibitem[Jimenez et~al.(2024)Jimenez, Yang, Wettig, Yao, Pei, Press, and Narasimhan]{jimenez2024swebenchlanguagemodelsresolve}
Carlos~E. Jimenez, John Yang, Alexander Wettig, Shunyu Yao, Kexin Pei, Ofir Press, and Karthik Narasimhan.
\newblock Swe-bench: Can language models resolve real-world github issues?, 2024.
\newblock URL \url{https://arxiv.org/abs/2310.06770}.

\bibitem[Rashid et~al.(2025)Rashid, Bock, Zhuang, Buchholz, Esler, Valentin, Franceschi, Wistuba, Sivaprasad, Kim, Deoras, Zappella, and Callot]{rashid2025swepolybenchmultilanguagebenchmarkrepository}
Muhammad~Shihab Rashid, Christian Bock, Yuan Zhuang, Alexander Buchholz, Tim Esler, Simon Valentin, Luca Franceschi, Martin Wistuba, Prabhu~Teja Sivaprasad, Woo~Jung Kim, Anoop Deoras, Giovanni Zappella, and Laurent Callot.
\newblock Swe-polybench: A multi-language benchmark for repository level evaluation of coding agents, 2025.
\newblock URL \url{https://arxiv.org/abs/2504.08703}.

\bibitem[Mathai et~al.(2024)Mathai, Huang, Maniatis, Nogikh, Ivan{\v{c}}i{\'c}, Yang, and Ray]{mathai2024kgym}
Alex Mathai, Chenxi Huang, Petros Maniatis, Aleksandr Nogikh, Franjo Ivan{\v{c}}i{\'c}, Junfeng Yang, and Baishakhi Ray.
\newblock Kgym: A platform and dataset to benchmark large language models on linux kernel crash resolution.
\newblock \emph{Advances in Neural Information Processing Systems}, 37:\penalty0 78053--78078, 2024.

\bibitem[Gu et~al.(2024)Gu, Rozière, Leather, Solar-Lezama, Synnaeve, and Wang]{gu2024cruxevalbenchmarkcodereasoning}
Alex Gu, Baptiste Rozière, Hugh Leather, Armando Solar-Lezama, Gabriel Synnaeve, and Sida~I. Wang.
\newblock Cruxeval: A benchmark for code reasoning, understanding and execution, 2024.
\newblock URL \url{https://arxiv.org/abs/2401.03065}.

\bibitem[Ding et~al.(2023{\natexlab{a}})Ding, Steenhoek, Pei, Kaiser, Le, and Ray]{ding2023tracedexecutionawarepretrainingsource}
Yangruibo Ding, Ben Steenhoek, Kexin Pei, Gail Kaiser, Wei Le, and Baishakhi Ray.
\newblock Traced: Execution-aware pre-training for source code, 2023{\natexlab{a}}.
\newblock URL \url{https://arxiv.org/abs/2306.07487}.

\bibitem[Ding et~al.(2024)Ding, Peng, Min, Kaiser, Yang, and Ray]{ding2024semcodertrainingcodelanguage}
Yangruibo Ding, Jinjun Peng, Marcus~J. Min, Gail Kaiser, Junfeng Yang, and Baishakhi Ray.
\newblock Semcoder: Training code language models with comprehensive semantics reasoning, 2024.
\newblock URL \url{https://arxiv.org/abs/2406.01006}.

\bibitem[Xu et~al.(2024)Xu, Cao, Lu, Lin, Han, He, Cheung, and Sun]{xu2024cruxeval}
Ruiyang Xu, Jialun Cao, Yaojie Lu, Hongyu Lin, Xianpei Han, Ben He, Shing-Chi Cheung, and Le~Sun.
\newblock Cruxeval-x: A benchmark for multilingual code reasoning, understanding and execution.
\newblock \emph{arXiv preprint arXiv:2408.13001}, 2024.

\bibitem[Chen et~al.(2024)Chen, Pan, Hu, Li, Li, and Xia]{chen2024reasoningruntimebehaviorprogram}
Junkai Chen, Zhiyuan Pan, Xing Hu, Zhenhao Li, Ge~Li, and Xin Xia.
\newblock Reasoning runtime behavior of a program with llm: How far are we?, 2024.
\newblock URL \url{https://arxiv.org/abs/2403.16437}.

\bibitem[Liu et~al.(2024)Liu, Zhang, Ibrahimzada, and Jabbarvand]{liu2024codemindframeworkchallengelarge}
Changshu Liu, Shizhuo~Dylan Zhang, Ali~Reza Ibrahimzada, and Reyhaneh Jabbarvand.
\newblock Codemind: A framework to challenge large language models for code reasoning, 2024.
\newblock URL \url{https://arxiv.org/abs/2402.09664}.

\bibitem[Xie et~al.(2025)Xie, Zheng, Liu, Wang, Wang, Tan, and Zhang]{xie2025corebenchmarkingllmscode}
Danning Xie, Mingwei Zheng, Xuwei Liu, Jiannan Wang, Chengpeng Wang, Lin Tan, and Xiangyu Zhang.
\newblock Core: Benchmarking llms code reasoning capabilities through static analysis tasks, 2025.
\newblock URL \url{https://arxiv.org/abs/2507.05269}.

\bibitem[Steenhoek et~al.(2025)Steenhoek, Rahman, Roy, Alam, Tong, Das, Barr, and Le]{steenhoek2025errmachinevulnerabilitydetection}
Benjamin Steenhoek, Md~Mahbubur Rahman, Monoshi~Kumar Roy, Mirza~Sanjida Alam, Hengbo Tong, Swarna Das, Earl~T. Barr, and Wei Le.
\newblock To err is machine: Vulnerability detection challenges llm reasoning, 2025.
\newblock URL \url{https://arxiv.org/abs/2403.17218}.

\bibitem[Anthropic(2024)]{anthropic2024claude35}
Anthropic.
\newblock Introducing claude 3.5 sonnet, June 2024.
\newblock URL \url{https://www.anthropic.com/news/claude-3-5-sonnet}.
\newblock Accessed: 2025-05-15.

\bibitem[OpenAI(2024)]{openai2024gpt4omini}
OpenAI.
\newblock Gpt-4o mini: Advancing cost-efficient intelligence, July 2024.
\newblock URL \url{https://openai.com/index/gpt-4o-mini-advancing-cost-efficient-intelligence/}.
\newblock Accessed: 2025-05-15.

\bibitem[Google(2025)]{google2025gemini15flash}
Google.
\newblock Gemini models: Gemini 1.5 flash, 2025.
\newblock URL \url{https://ai.google.dev/gemini-api/docs/models#gemini-1.5-flash}.
\newblock Accessed: 2025-05-15.

\bibitem[Allamanis et~al.(2021)Allamanis, Jackson-Flux, and Brockschmidt]{allamanis2021self}
Miltiadis Allamanis, Henry Jackson-Flux, and Marc Brockschmidt.
\newblock Self-supervised bug detection and repair.
\newblock In \emph{NeurIPS}, 2021.

\bibitem[Krekel et~al.(2004)Krekel, Oliveira, Pfannschmidt, Bruynooghe, Laugher, and Bruhin]{pytest}
Holger Krekel, Bruno Oliveira, Ronny Pfannschmidt, Floris Bruynooghe, Brianna Laugher, and Florian Bruhin.
\newblock pytest x.y.
\newblock \url{https://github.com/pytest-dev/pytest}, 2004.
\newblock Version x.y. Contributors include Holger Krekel, Bruno Oliveira, Ronny Pfannschmidt, Floris Bruynooghe, Brianna Laugher, Florian Bruhin, and others.

\bibitem[Rachum et~al.(2019)Rachum, Hall, Yanokura, et~al.]{rachum2019pysnooper}
Ram Rachum, Alex Hall, Iori Yanokura, et~al.
\newblock Pysnooper: Never use print for debugging again, jun 2019.
\newblock URL \url{https://github.com/cool-RR/PySnooper}.

\bibitem[Arya et~al.(2023)Arya, Chang, Metzman, Serebryany, and Liu]{ossfuzz}
Abhishek Arya, Oliver Chang, Jonathan Metzman, Kostya Serebryany, and Dongge Liu.
\newblock Oss-fuzz, 2023.
\newblock URL \url{https://github.com/google/oss-fuzz}.
\newblock Accessed: 2025-05-14.

\bibitem[{Free Software Foundation}()]{gdb}
{Free Software Foundation}.
\newblock {GDB}: The {GNU} project debugger.
\newblock \url{https://www.sourceware.org/gdb/}.
\newblock Accessed: May 15, 2025.

\bibitem[Fraser and Arcuri(2012)]{ICSE12}
Gordon Fraser and Andrea Arcuri.
\newblock Sound empirical evidence in software testing.
\newblock In \emph{34th International Conference on Software Engineering, ICSE 2012, June 2-9, 2012, Zurich, Switzerland}, pages 178--188. IEEE, 2012.
\newblock ISBN 978-1-4673-1067-3.

\bibitem[Fraser and Arcuri(2011)]{10.1145/2025113.2025179}
Gordon Fraser and Andrea Arcuri.
\newblock Evosuite: automatic test suite generation for object-oriented software.
\newblock In \emph{Proceedings of the 19th ACM SIGSOFT Symposium and the 13th European Conference on Foundations of Software Engineering}, ESEC/FSE '11, page 416–419, New York, NY, USA, 2011. Association for Computing Machinery.
\newblock ISBN 9781450304436.
\newblock \doi{10.1145/2025113.2025179}.
\newblock URL \url{https://doi.org/10.1145/2025113.2025179}.

\bibitem[{Oracle Corporation}(2025)]{oracleJDB}
{Oracle Corporation}.
\newblock \emph{JDB - The Java Debugger}, 2025.
\newblock URL \url{https://docs.oracle.com/javase/8/docs/technotes/tools/unix/jdb.html}.
\newblock Part of the Java SE Development Kit.

\bibitem[Chen et~al.(2021{\natexlab{a}})Chen, Tworek, Jun, Yuan, de~Oliveira~Pinto, Kaplan, Edwards, Burda, Joseph, Brockman, Ray, Puri, Krueger, Petrov, Khlaaf, Sastry, Mishkin, Chan, Gray, Ryder, Pavlov, Power, Kaiser, Bavarian, Winter, Tillet, Such, Cummings, Plappert, Chantzis, Barnes, Herbert-Voss, Guss, Nichol, Paino, Tezak, Tang, Babuschkin, Balaji, Jain, Saunders, Hesse, Carr, Leike, Achiam, Misra, Morikawa, Radford, Knight, Brundage, Murati, Mayer, Welinder, McGrew, Amodei, McCandlish, Sutskever, and Zaremba]{chen2021evaluatinglargelanguagemodels}
Mark Chen, Jerry Tworek, Heewoo Jun, Qiming Yuan, Henrique~Ponde de~Oliveira~Pinto, Jared Kaplan, Harri Edwards, Yuri Burda, Nicholas Joseph, Greg Brockman, Alex Ray, Raul Puri, Gretchen Krueger, Michael Petrov, Heidy Khlaaf, Girish Sastry, Pamela Mishkin, Brooke Chan, Scott Gray, Nick Ryder, Mikhail Pavlov, Alethea Power, Lukasz Kaiser, Mohammad Bavarian, Clemens Winter, Philippe Tillet, Felipe~Petroski Such, Dave Cummings, Matthias Plappert, Fotios Chantzis, Elizabeth Barnes, Ariel Herbert-Voss, William~Hebgen Guss, Alex Nichol, Alex Paino, Nikolas Tezak, Jie Tang, Igor Babuschkin, Suchir Balaji, Shantanu Jain, William Saunders, Christopher Hesse, Andrew~N. Carr, Jan Leike, Josh Achiam, Vedant Misra, Evan Morikawa, Alec Radford, Matthew Knight, Miles Brundage, Mira Murati, Katie Mayer, Peter Welinder, Bob McGrew, Dario Amodei, Sam McCandlish, Ilya Sutskever, and Wojciech Zaremba.
\newblock Evaluating large language models trained on code, 2021{\natexlab{a}}.
\newblock URL \url{https://arxiv.org/abs/2107.03374}.

\bibitem[Chen et~al.(2021{\natexlab{b}})Chen, Tworek, Jun, Yuan, Pinto, Kaplan, Edwards, Burda, Joseph, Brockman, et~al.]{chen2021evaluating}
Mark Chen, Jerry Tworek, Heewoo Jun, Qiming Yuan, Henrique Ponde De~Oliveira Pinto, Jared Kaplan, Harri Edwards, Yuri Burda, Nicholas Joseph, Greg Brockman, et~al.
\newblock Evaluating large language models trained on code.
\newblock \emph{arXiv preprint arXiv:2107.03374}, 2021{\natexlab{b}}.

\bibitem[Austin et~al.(2021)Austin, Odena, Nye, Bosma, Michalewski, Dohan, Jiang, Cai, Terry, Le, et~al.]{austin2021program}
Jacob Austin, Augustus Odena, Maxwell Nye, Maarten Bosma, Henryk Michalewski, David Dohan, Ellen Jiang, Carrie Cai, Michael Terry, Quoc Le, et~al.
\newblock Program synthesis with large language models.
\newblock \emph{arXiv preprint arXiv:2108.07732}, 2021.

\bibitem[Puri et~al.(2021)Puri, Kung, Janssen, Zhang, Domeniconi, Zolotov, Dolby, Chen, Choudhury, Decker, et~al.]{puri2021codenet}
Ruchir Puri, David~S Kung, Geert Janssen, Wei Zhang, Giacomo Domeniconi, Vladimir Zolotov, Julian Dolby, Jie Chen, Mihir Choudhury, Lindsey Decker, et~al.
\newblock Codenet: A large-scale ai for code dataset for learning a diversity of coding tasks.
\newblock \emph{arXiv preprint arXiv:2105.12655}, 2021.

\bibitem[Zhang et~al.(2024{\natexlab{a}})Zhang, Zhang, Ran, Zhu, Dou, Hao, Xie, and Zhang]{Zhang_2024}
Yakun Zhang, Wenjie Zhang, Dezhi Ran, Qihao Zhu, Chengfeng Dou, Dan Hao, Tao Xie, and Lu~Zhang.
\newblock Learning-based widget matching for migrating gui test cases.
\newblock In \emph{Proceedings of the IEEE/ACM 46th International Conference on Software Engineering}, ICSE ’24. ACM, February 2024{\natexlab{a}}.
\newblock \doi{10.1145/3597503.3623322}.
\newblock URL \url{http://dx.doi.org/10.1145/3597503.3623322}.

\bibitem[Li et~al.(2024)Li, Li, Zhang, Dong, and Jin]{li2024evocodebenchevolvingcodegeneration}
Jia Li, Ge~Li, Xuanming Zhang, Yihong Dong, and Zhi Jin.
\newblock Evocodebench: An evolving code generation benchmark aligned with real-world code repositories, 2024.
\newblock URL \url{https://arxiv.org/abs/2404.00599}.

\bibitem[Zhang et~al.(2024{\natexlab{b}})Zhang, Li, Li, Shi, and Jin]{zhang2024codeagentenhancingcodegeneration}
Kechi Zhang, Jia Li, Ge~Li, Xianjie Shi, and Zhi Jin.
\newblock Codeagent: Enhancing code generation with tool-integrated agent systems for real-world repo-level coding challenges, 2024{\natexlab{b}}.
\newblock URL \url{https://arxiv.org/abs/2401.07339}.

\bibitem[Yu et~al.(2024)Yu, Shen, Ran, Zhang, Zhang, Ma, Liang, Li, Wang, and Xie]{10.1145/3597503.3623316}
Hao Yu, Bo~Shen, Dezhi Ran, Jiaxin Zhang, Qi~Zhang, Yuchi Ma, Guangtai Liang, Ying Li, Qianxiang Wang, and Tao Xie.
\newblock Codereval: A benchmark of pragmatic code generation with generative pre-trained models.
\newblock In \emph{Proceedings of the IEEE/ACM 46th International Conference on Software Engineering}, ICSE '24, New York, NY, USA, 2024. Association for Computing Machinery.
\newblock ISBN 9798400702174.
\newblock \doi{10.1145/3597503.3623316}.
\newblock URL \url{https://doi.org/10.1145/3597503.3623316}.

\bibitem[Chen et~al.(2025)Chen, Pusarla, and Ray]{chen2025dynamic}
Simin Chen, Pranav Pusarla, and Baishakhi Ray.
\newblock Dynamic benchmarking of reasoning capabilities in code large language models under data contamination.
\newblock \emph{arXiv preprint arXiv:2503.04149}, 2025.

\bibitem[Du et~al.(2023)Du, Liu, Wang, Wang, Liu, Chen, Feng, Sha, Peng, and Lou]{du2023classeval}
Xueying Du, Mingwei Liu, Kaixin Wang, Hanlin Wang, Junwei Liu, Yixuan Chen, Jiayi Feng, Chaofeng Sha, Xin Peng, and Yiling Lou.
\newblock Classeval: A manually-crafted benchmark for evaluating llms on class-level code generation.
\newblock \emph{arXiv preprint arXiv:2308.01861}, 2023.

\bibitem[Izadi et~al.(2024)Izadi, Katzy, Van~Dam, Otten, Popescu, and Van~Deursen]{10.1145/3597503.3639138}
Maliheh Izadi, Jonathan Katzy, Tim Van~Dam, Marc Otten, Razvan~Mihai Popescu, and Arie Van~Deursen.
\newblock Language models for code completion: A practical evaluation.
\newblock In \emph{Proceedings of the IEEE/ACM 46th International Conference on Software Engineering}, ICSE '24, New York, NY, USA, 2024. Association for Computing Machinery.
\newblock ISBN 9798400702174.
\newblock \doi{10.1145/3597503.3639138}.
\newblock URL \url{https://doi.org/10.1145/3597503.3639138}.

\bibitem[Ding et~al.(2023{\natexlab{b}})Ding, Wang, Ahmad, Ding, Tan, Jain, Ramanathan, Nallapati, Bhatia, Roth, and Xiang]{ding2023crosscodeevaldiversemultilingualbenchmark}
Yangruibo Ding, Zijian Wang, Wasi~Uddin Ahmad, Hantian Ding, Ming Tan, Nihal Jain, Murali~Krishna Ramanathan, Ramesh Nallapati, Parminder Bhatia, Dan Roth, and Bing Xiang.
\newblock Crosscodeeval: A diverse and multilingual benchmark for cross-file code completion, 2023{\natexlab{b}}.
\newblock URL \url{https://arxiv.org/abs/2310.11248}.

\bibitem[Kwon et~al.(2023)Kwon, Li, Zhuang, Sheng, Zheng, Yu, Gonzalez, Zhang, and Stoica]{kwon2023efficient}
Woosuk Kwon, Zhuohan Li, Siyuan Zhuang, Ying Sheng, Lianmin Zheng, Cody~Hao Yu, Joseph~E. Gonzalez, Hao Zhang, and Ion Stoica.
\newblock Efficient memory management for large language model serving with pagedattention.
\newblock In \emph{Proceedings of the ACM SIGOPS 29th Symposium on Operating Systems Principles}, 2023.

\bibitem[Yan et~al.(2024)Yan, Liu, Wang, Li, Chen, Wang, Lin, Zhao, Zhu, Sundaram, and Deng]{yan2024codescopeexecutionbasedmultilingualmultitask}
Weixiang Yan, Haitian Liu, Yunkun Wang, Yunzhe Li, Qian Chen, Wen Wang, Tingyu Lin, Weishan Zhao, Li~Zhu, Hari Sundaram, and Shuiguang Deng.
\newblock Codescope: An execution-based multilingual multitask multidimensional benchmark for evaluating llms on code understanding and generation, 2024.
\newblock URL \url{https://arxiv.org/abs/2311.08588}.

\end{thebibliography}

%\newpage
%\input{Text/checklist}

%%%%%%%%%%%%%%%%%%%%%%%%%%%%%%%%%%%%%%%%%%%%%%%%%%%%%%%%%%%%

\newpage
\appendix

% \begin{document}
\section{Appendix}
\label{appendix}

% \subsection{Model Name Mappings for Figure Labels}

% \begin{table}[h]
% \centering
% \caption{Model Names and Their IDs in the figures }\label{tab:model-names}
% \centering
% \footnotesize
% \setlength{\tabcolsep}{1pt}
%   \resizebox{0.6\textwidth}{!}{
% \begin{tabular}{@{}ll@{}}
% \toprule
% \textbf{Full Model Name} & \textbf{Model ID in the figures}\\
% \midrule
% openai/gpt-4.0-mini & GPT-4o (Reasoning) \\
% anthropic.claude-3-5-sonnet-20241022-v2:0 & CL-3.5 (Reasoning) \\
% gemini-1.5-flash-002 & Gem-1.5 (Reasoning) \\
% meta-llama/Llama-3.1-8B-Instruct & L-3.1\\
% Qwen/Qwen2.5-14B-Instruct-1M & Q-2.5\\
% Qwen/Qwen2.5-Coder-7B-Instruct & Qwen2.5-C\\
% deepseek-ai/DeepSeek-Coder-V2-Lite-Instruct & DS-C\\
% microsoft/Phi-4-mini-instruct & Phi-4\\
% microsoft/Phi-3.5-mini-instruct & Phi-3.5\\
% ibm-granite/granite-3.2-8b-instruct & Gr-3.2 (Reasoning)\\
% deepseek-ai/DeepSeek-R1-Distill-Qwen-7B & DSR1-Q-7B (Reasoning) \\
% deepseek-ai/DeepSeek-R1-Distill-Llama-8B & DSR1-L (Reasoning)\\
% deepseek-ai/DeepSeek-R1-Distill-Qwen-14B & DSR1-Q-14B  (Reasoning)\\
% ibm-granite/granite-3.2-8b-instruct-preview & Granite-3.2 Pr (Reasoning)\\

% \bottomrule
% \end{tabular}
% }
% \end{table}

\subsection{Computation Resources and Inference Tools}
\label{app:computation}
All our experiments were conducted using the following computational resources:

\begin{itemize}
    \item \textbf{GPU:} NVIDIA RTX A6000 with 49GB VRAM
    \item \textbf{Memory Utilization:} ~0.9GB GPU memory during inference runs
    \item \textbf{Software Stack:} vLLM ~\citep{kwon2023efficient} for optimised transformer inference
    \item \textbf{Operations:} Inference-only experiments (no fine-tuning performed)
\end{itemize}

The inference parameters were controlled through the following configuration \ref{tab:config}:

\begin{table}[H]
\centering
\caption{Inference Configuration Parameters}
\label{tab:config}
\begin{tabular}{@{}llp{6cm}@{}}
\toprule
\textbf{Parameter} & \textbf{Value} & \textbf{Description} \\ \midrule
temperature & 0.8 & Controls randomness: Lower = more deterministic \\ 
top\_p & 0.95 & Nucleus sampling: Only top 95\% probability mass \\
max\_tokens & 4096 (default) & Base context window size \\
 & 16384 (For reasoning models) & Extended context for specific models \\
tp\_size & 1 & No tensor parallelism \\
dtype & float16 & Half-precision floating point \\
stop & \texttt{[\textbackslash n>>>}, \texttt{\textbackslash n\$}, ...] & Generation stopping tokens \\ \bottomrule
\end{tabular}
\end{table}

\subsection{Limitations and Future Work}
\label{app:limitations}
In this work, we consider exact matching between the model's response and ground truth as correct. In future work, we would like to explore other metrics such as pass@k. However, we did explore models' performance of computing abstract value prediction/approximation of code semantics.

%That's why the model's performance on concrete value prediction is quite low, however, this problem is mitigated by abstract value prediction/approximation of code semantics. Still, we would like to explore other metrics for concrete value prediction tasks.

For some RQs, we only evaluated the models on subset of tasks. For example, when comparing different programming languages in RQ6, we used input/output predictions. In the future, we will extend such evaluations to more tasks. When computing pointer alias in RQ3, we used C languages. Future work can also include object aliasing detection for Python and Java. 

Additionally, we would like to expand our framework to support project tracing and task-specific benchmark datasets beyond the scope of three languages. This includes adding other languages with diverse domains like (e.g., functional, system language), to enable a more comprehensive evaluation of LLMs.

In future, it will be also be interesting to explore more advanced prompting techniques and even fine-tune the models to further evaluate them.
%For this work, we didn't curate all task-specific datasets across all three languages. For example, we only curated a pointer aliasing dataset on C, while analogous object aliasing datasets for Python and Java remain unexplored. This limits our ability to generalise findings with languages that can use memory directly, like C, and languages with abstraction (Python and Java).\\

\subsection{Trace Collection}
In this section, we will discuss the detailed process to generate the C, Python and Java real-world traces.

\subsubsection{Real World Project Collection}
\begin{enumerate}
        \item \textbf{C Project Collection}: For our research, we wanted to generate trace dataset on real-world projects. We have selected real-world C projects that were curated in the OSS-FUZZ repository. The primary reason behind selecting these projects is that they represent different domains to ensure diversity in software types. This choice also aligns with our research goal to generate a benchmark real-world C trace dataset. 
        \item \textbf{Python Project Collection}: To collect real-world Python projects, we adopted two approaches. 1) We cloned all 1489 repositories from GitHub that appear in the PyPIBugs dataset, which was released in 2021 \cite{allamanis2021self}. 2) To avoid missing popular projects after 2021, we use GitHub API to search for repositories that are marked as mainly written in Python and get the results according to the descending order of the number of stars. To maximise the probability that we can execute them easily with the pytest module, we only consider projects that seemingly have a testing folder at the top or second level. For better compatibility and the reflection of the recent trend of programming styles, we further filtered out projects that have not been updated in the last four years. Finally, we got 544 projects.
        \item \textbf{Java Dataset Collection}: For Java, we aimed to gather a diverse set of real-world projects to have a comprehensive trace analysis, which aligns with our trace dataset generation objective. We have used EvoSuite \cite{10.1145/2025113.2025179} for test suite generation, and the SF110 dataset has been used as it is recommended by EvoSuite. This choice ensures compatibility and a high testing coverage rate.
        
\end{enumerate}

\subsubsection{High-Level Steps Overview of Generating Traces}
\paragraph{C Trace Collection}
\begin{enumerate}
    \item \textbf{Building Projects}: Building projects before fuzzing is necessary to ensure that different project dependencies are correctly installed and configured, avoiding runtime errors during the fuzzing process. This helps to create a consistent and effective environment for the next fuzzing process.
    \item \textbf{Fuzzing}: In the fuzzing phase, we executed the fuzzer on the already-built projects to generate the input data corpora. We configure the fuzzing tools with appropriate settings and parameters for each project. This includes specifying input seed files, maximum time for the fuzzer to run and kill delay to maximise code coverage. Throughout the fuzzing process, detailed logs are maintained to track the execution progress and other relevant information. These logs can aid in debugging, result interpretation, and fuzzing outcomes.
    \item \textbf{Tracing}: Tracing is the most crucial step to have the execution information of real-world projects. We use a tracing framework with the GNU debugger to log the execution of the projects. With the help of the framework, we log function calls, variable values, and other states during the execution of the projects. We start the tracing by setting an entry point for the program, and during the execution of the tracing, we record the different states of the program at various points by logging them into an XML-formatted file for further analysis. Additionally, we have added a tracer timeout to ensure the maximum running time of the tracer, as well as an extra kill delay to ensure the safe exit of the tracer. This ensures the reliability and robustness of the tracing process if any unexpected events occur.
\end{enumerate}

\paragraph{Python Trace Collection}

\begin{enumerate}
    \item \textbf{Execution}: We execute the collected projects to get traces in a best-effort approach: 1) We scan the common dependency files to install the dependencies into an independent Python environment for each project. 2) We use pytest to execute the test cases in the projects and collect the outputs. 3) We analyze the outputs to identify missing dependency errors and try to install the missing dependencies several times.
    \item \textbf{Tracing}: We use the PySnooper tool to trace the projects but made the following modifications to it: 1) We only keep traces corresponding to source code files in the project source directories to exclude traces happened in Python built-in functions or third-party dependencies. 2) We expand the representation of user-defined class objects by showing the name and value pairs of their first-level attributes. 3) We save the types of variables in traces instead of just value representations to provide more information for the execution-aware source code modeling.
\end{enumerate}

\newpage
\paragraph{Java Trace Collection}
\begin{enumerate}
    \item \textbf{Tracing}: To generate the tracing framework for Java, we integrated Java Debugger to record the execution details of the projects. We logged method invocation, variable values, and different program states during the execution cycle. For Java, we stored the raw trace in JSON format, which was stored in directories specific to each class within the project directories. This helps manage large amounts of data, consequently making it easier to retrieve, analyze and clean it up for further tasks.  
\end{enumerate}
\begin{table}[htbp]
  \centering
  \renewcommand{\arraystretch}{1}
  \caption{Trace Collection}
  \label{tab:components-simple}
  \begin{tabular}{lccc}
    \toprule
    \toprule
    \textbf{} & \textbf{Real-world} & \textbf{Testing} & \textbf{Tracing} \\
    \textbf{} & \textbf{projects} & \textbf{Tools} & \textbf{Tools} \\
    \midrule
    Python & PyPIbugs+Github (544) & Pytest & Pysnooper \\
    C & OSS-Fuzz (100) & Fuzzing & GDB \\
    Java & SF-110 (100) & Evosuite & JDB \\
    \bottomrule
    \bottomrule
  \end{tabular}
\end{table}

\subsection{Language Selection Rationale and Extensibility} \label{app:language_rationale}
The three programming languages in our benchmark are important and representative for programming language features and real-world applications: C is a low level programming language and useful for building systems, Java has object-oriented programming features, are widely used for building enterprise and web applications, Python is important for data science and AI applications. 
Our benchmark is extensible, third-parties (as well as ourselves in future) can add more languages. Our scripts, prompts and methodologies can be adapted for new programming languages to (1) select and download programs of a programming language from GitHub, (2) fuzz for generating test inputs, (3) trace and curate ground truth data (4) provide input and parse output when interacting with models. Instead of GDB (for C), Pysnooper (for Python) and Java Debugger Oracle Corporation (for Java), we will need to plug in debugging tools for new programming languages.

You may ask “we have compilers and code execution tools, why do we need models to predict dynamic information”---- Predicting dynamic values is not only useful for executing programs, but required for many other downstream tasks. For example, in \figref{fig:example}, to generate test input that can exercise a true branch, the models need to know how each operator in statements updates the values. The key difference is that: when using code execution tools, we give one input and ask for the output, but in other downstream tasks, we require models to first understand fine-grained semantics and then find inputs that can satisfy certain constraints. 
Even for compiling and executing programs, we may benefit from LLMs prediction, as compilation and execution can be time-consuming and hard to be configured, especially for legacy code. This ability of LLMs is particularly important when the user cannot run the code snippet, e.g., missing dependencies or unavailable resources.
Predicting input/output values as code reasoning tasks have been established by prior research \citep{gu2024cruxevalbenchmarkcodereasoning}, \citep{chen2021evaluating}, \citep{yan2024codescopeexecutionbasedmultilingualmultitask}.  Our work extended prior research by introducing real-world projects and fine-grained tasks supported only by our tracing framework.

\subsection{Detailed Description of Tasks}
In this section, we will provide a detailed description of our tasks. \ref{statement-task} provides a comprehensive description of the statement-based evaluation we performed on LLMs. We sampled five types of statements, i.e, Assignment, Arithmetic, Constant, Boolean, and Function Call, and prompted the models about the value after execution of each type of statement given the variable states before executing that statement. For the block prediction task \ref{block-task}, we sampled statements from the start of the code snippet, given the input of the code, we prompted the model to predict the output at the end of the 1st statement, the 2nd  statement, and the 3rd statement. For Branch prediction \ref{branch-task}, we promoted the model whether a specific branch will be taken or not of a code snippet, given the input of that code snippet. 

In the case of \ref{loop-task}, we sampled loop statements from the code snippets. For each loop, we first collected the number of iterations of that loop as ground truth, and queried the model about how many times the loop would be iterated. We also collected and prompted the model regarding the variable state inside the loop body after the n-th interaction. We have named this "In-Loop" prediction. Additionally, we sampled variables after the execution of the whole loop and queried the model regarding the variable value after the execution of the loop body ("Post-Loop" prediction). For input/output prediction \ref{io-task}, we used the approach similar to ~\cite{gu2024cruxevalbenchmarkcodereasoning}. For output prediction, we give the entire code snippet and the input of the code snippet, and vice versa for input prediction.

\scalebox{0.9}{
\begin{minipage}{1.05\textwidth}

\subsubsection{Statement Prediction Task}
\label{statement-task}
\begin{tcolorbox}[colback=white,colframe=black!75!white]
\begin{lstlisting}[language=Python]
def xldate_from_date_tuple(date_tuple, datemode):
    year, month, day = date_tuple
    data_list = list(data_tuple)
    if datemode not in (0, 1):
        raise XLDateBadDatemode(datemode)

    if year == 0 and month == 0 and day == 0:
        return 0.00
    c = 100
    if not (1900 <= year <= 9999):
        raise XLDateBadTuple("Invalid_year:%r" % ((year, month, day),))
    if not (1 <= month <= 12):
        raise XLDateBadTuple("Invalid_month:%r" % ((year, month, day),))
    if day < 1 \
    or (day > _days_in_month[month] and not (day == 29 and month == 2 and _leap(year))):
        raise XLDateBadTuple("Invalid_day:%r" % ((year, month, day),))

    Yp = year + 4716
    M = month
    if M <= 2:
        Yp = Yp - 1
        Mp = M + 9
    else:
        Mp = M - 3
    jdn = ifd(1461 * Yp, 4) + ifd(979 * Mp + 16, 32) + \
        day - 1364 - ifd(ifd(Yp + 184, 100) * 3, 4)
    xldays = jdn - _JDN_delta[datemode]
    if xldays <= 0:
        raise XLDateBadTuple("Invalid(year,month,day):%r" % ((year, month, day),))
    if xldays < 61 and datemode == 0:
        raise XLDateAmbiguous("Before1900-03-01:%r" % ((year, month, day),))
    return float(xldays)

xldate_from_date_tuple(date_tuple=(1907, 7, 3), datemode=0)
\end{lstlisting}

\vspace{10pt}

\noindent\textbf{Assignment Prediction}
\begin{itemize}
    \item What will be the value of the final output of the statement \texttt{year, month, day = date\_tuple} given \texttt{\{'date\_tuple': (1907, 7, 3)\}} after executing the statement?
\end{itemize}

\vspace{5pt}
\noindent\textbf{Arithmetic Prediction}
\begin{itemize}
    \item What will be the value of the final output of the statement \texttt{Yp = year + 4716} given \texttt{\{'year': 1907\}} after executing the statement?
\end{itemize}

\vspace{5pt}
\noindent\textbf{Constant Prediction}
\begin{itemize}
    \item What will be the value of the final output of the statement \texttt{c = 0} given \texttt{\{'constant': 0\}} after executing the statement?
\end{itemize}

\vspace{5pt}
\noindent\textbf{Boolean Prediction}
\begin{itemize}
    \item Will the true branch of the statement \texttt{if datemode not in (0, 1):} be executed given \texttt{\{'datemode': 0\}}?
\end{itemize}

\vspace{5pt}
\noindent\textbf{Function Call Prediction}
\begin{itemize}
    \item What will be the value of the final output of the statement \texttt{data\_list = list(data\_tuple)} given \texttt{\{'date\_tuple': (1907, 7, 3)\}} after executing the statement??
\end{itemize}

\end{tcolorbox}

\end{minipage}
}

\scalebox{0.9}{
\begin{minipage}{1.05\textwidth}
\subsubsection{Block Prediction Task}
\label{block-task}
\begin{tcolorbox}[colback=white,colframe=black!75!white]
\begin{lstlisting}[language=Python]
def exchange(a, i, j):
    temp = a[i]
    a[i] = a[j]
    a[j] = temp
exchange(a=[0, 100, 200, 0, 0, 0, 0, 0, 0, 0], i=2, j=1)
\end{lstlisting}

\vspace{10pt}

\noindent\textbf{1-Block Prediction}
\begin{itemize}
    \item What will be the value of the statement \texttt{temp = a[i]} given the function input \texttt{{'a': [0, 100, 200, 0, 0, 0, 0, 0, 0, 0], 'i': 2, 'j': 1}}?
\end{itemize}

\noindent\textbf{2-Block Prediction}
\begin{itemize}
    \item What will be the value of the statement \texttt{a[i] = a[j]} given the function input \texttt{{'a': [0, 100, 200, 0, 0, 0, 0, 0, 0, 0], 'i': 2, 'j': 1}}?
\end{itemize}

\noindent\textbf{3-Block Prediction}
\begin{itemize}
    \item What will be the value of the statement \texttt{a[j] = temp} given the function input \texttt{{'a': [0, 100, 200, 0, 0, 0, 0, 0, 0, 0], 'i': 2, 'j': 1}}?
\end{itemize}

\end{tcolorbox}

\subsubsection{Branch Task}
\label{branch-task}
\begin{tcolorbox}[colback=white,colframe=black!75!white]
\begin{lstlisting}[language=Python]
1.def xldate_from_date_tuple(date_tuple, datemode):
2.   
3.    year, month, day = date_tuple
4.
5.    if datemode not in (0, 1):
6.        raise XLDateBadDatemode(datemode)
7.
8.    if year == 0 and month == 0 and day == 0:
9.        return 0.00
10.
11.    if not (1900 <= year <= 9999):
12.        raise XLDateBadTuple("Invalid_year:%r" % ((year, month, day),))
13.    if not (1 <= month <= 12):
14.        raise XLDateBadTuple("Invalid_month:%r" % ((year, month, day),))
15.    if  day < 1 \
16.    or (day > _days_in_month[month] and not(day == 29 and month == 2 and _leap(year))):
17.        raise XLDateBadTuple("Invalid_day:%r" % ((year, month, day),))
18.
19.    Yp = year + 4716
20.    M = month
21.    if M <= 2:
22.        Yp = Yp - 1
23.        Mp = M + 9
24.    else:
25.        Mp = M - 3
26.    jdn = ifd(1461 * Yp, 4) + ifd(979 * Mp + 16, 32) + \
27.        day - 1364 - ifd(ifd(Yp + 184, 100) * 3, 4)
28.    xldays = jdn - _JDN_delta[datemode]
29.    if xldays <= 0:
30.        raise XLDateBadTuple("Invalid(year,month,day):%r" % ((year, month, day),))
31.    if xldays < 61 and datemode == 0:
32.        raise XLDateAmbiguous("Before1900-03-01:%r" % ((year, month, day),))
33.    return float(xldays)
34.
35.xldate_from_date_tuple((1907, 7, 3), 0)
\end{lstlisting}

\vspace{10pt}

\noindent\textbf{Brach Prediction}
\begin{itemize}
    \item Is line 12, \texttt{raise XLDateBadTuple("Invalid year: \%r" \% ((year, month, day),))} executed when \texttt{xldate\_from\_date\_tuple((1907, 7, 3), 0)} is called?
\end{itemize}

\end{tcolorbox}

\end{minipage}
}

\subsubsection{Loop Task}
\label{loop-task}
\begin{tcolorbox}[colback=white,colframe=black!75!white]
\begin{lstlisting}[language=Python]
1. def make_version_tuple(vstr=None):
2.     if vstr is None:
3.         vstr = __version__
4.     if vstr[0] == "v":
5.         vstr = vstr[1:]
6.     components = []
7.     for component in vstr.split("+")[0].split("."):
8.         try:
9.             components.append(int(component))
10.         except ValueError:
11.             break
12.     components = tuple(components)
13.     return components
14.
15. make_version_tuple('v0.1.1')
\end{lstlisting}

\vspace{10pt}

\noindent\textbf{Iteration Prediction}
\begin{itemize}
    \item How many times will the loop on line 7 execute when  \texttt{make\_version\_tuple('v0.1.1')} is called?
\end{itemize}

\vspace{4pt}  % Reduced spacing between sections

\noindent\textbf{In-Loop Prediction}
\begin{itemize}
    \item What is the value of \texttt{components} in line 9 after 2nd iteration when  \texttt{make\_version\_tuple('v0.1.1')} is called?
\end{itemize}

\vspace{4pt}  % Consistent spacing between sections

\noindent\textbf{Post-Loop Prediction}
\begin{itemize}
    \item What is the value of \texttt{components} in line 12 when \texttt{make\_version\_tuple('v0.1.1')} is called?
\end{itemize}

\end{tcolorbox}

\subsubsection{Input-Output Task}
\label{io-task}
\begin{tcolorbox}[colback=white,colframe=black!75!white]
\begin{lstlisting}[language=Python]
def cast_tuple(val, length = None):
    if isinstance(val, list):
        val = tuple(val)

    output = val if isinstance(val, tuple) else ((val,) * default(length, 1))

    if exists(length):
        assert len(output) == length

    return output
\end{lstlisting}

\vspace{10pt}

\noindent\textbf{Output Prediction}
\begin{itemize}
    \item What will be the output of the code given input \texttt{\{'val':1, 'length':4\}}?
\end{itemize}

\vspace{4pt}  % Reduced spacing between sections

\noindent\textbf{Input Prediction}
\begin{itemize}
    \item What will be the input of the code given output \texttt{(1, 1, 1, 1)}?
\end{itemize}

\end{tcolorbox}

\subsection{Concrete to Abstract Mapping}
For the approximation of code semantics, we prompted the models to reason in abstract values, instead of reasoning about the concrete exact value. Table \ref{abstracttable} shows the mapping from concrete value to abstract category, following the prior literature~\cite{ding2023tracedexecutionawarepretrainingsource}.  When defining these mappings, we carefully aligned the value ranges for each abstract category with the overall value distribution observed in our benchmark. We evaluated the abstract mapping results against a random baseline, where mapping rules were selected randomly from all available mapping categories.
\begin{table}[htbp]
\centering
\renewcommand{\arraystretch}{1.2}
\caption{Concrete Value to Quantize Value Mapping}
\label{abstracttable}
\label{tab:value-types}
\begin{tabular}{|l|l|l|}
\hline
\textbf{Type} & \textbf{Condition} & \textbf{Category} \\
\hline
\multirow{6}{*}{Integer} 
 & $0 < v \leq 10$ & Positive Regular \\
 & $v > 10$ & Positive Large \\
 & $v == 0$ & Zero \\
 & $-10 \leq v < 0$ & Negative Regular \\
 & $v < -10$ & Negative Large \\
\hline
\multirow{7}{*}{Float} 
 & $1.0 < v \leq 10.0$ & Positive Regular \\
 & $0.0 < v \leq 1.0$ & Positive Small \\
 & $10.0 < v$ & Positive Large \\
 & $v == 0.0$ & Zero \\
 & $-1.0 \leq v < 0.0$ & Negative Small \\
 & $-10.0 \leq v < -1.0$ & Negative Regular \\
 & $v < -10.0$ & Negative Large \\
\hline
\multirow{4}{*}{String} 
 & $\texttt{len(s) == 0}$ & Empty String \\
 & $\texttt{len(s) > 0 and s.isalpha()}$ & Alphabetic String \\
 & $\texttt{len(s) > 0 and s.isdigit()}$ & Numeric String \\
 & $\texttt{len(s) > 0 and not (s.isalpha() or s.isdigit())}$ & Mixed String \\
\hline
\multirow{2}{*}{List} 
 & $\texttt{len(lst) == 0}$ & Empty List \\
 & $\texttt{len(lst) > 0}$ & Non-Empty List \\
\hline
\multirow{2}{*}{Tuple} 
 & $\texttt{len(tup) == 0}$ & Empty Tuple \\
 & $\texttt{len(tup) > 0}$ & Non-Empty Tuple \\
\hline
\multirow{2}{*}{Dict} 
 & $\texttt{len(dict) == 0}$ & Empty Dictionary \\
 & $\texttt{len(dict) > 0}$ & Non-Empty Dictionary \\
\hline
\multirow{2}{*}{Set} 
 & $\texttt{len(set) == 0}$ & Empty Set \\
 & $\texttt{len(set) > 0}$ & Non-Empty Set \\
\hline
\multirow{2}{*}{Boolean} 
 & $\texttt{True}$ & True \\
 & $\texttt{False}$ & False \\
\hline
NoneType & \texttt{None} & None \\
\hline
\end{tabular}
\end{table}

\newpage
\subsection{Prompting Techniques}
The following is a subset of prompts we used to evaluate the models.  The rest prompts are shown in our data package.

\newcommand{\promptbox}[2]{
\begin{tcolorbox}[title=#1, colback=blue!5!white, colframe=blue!75!black]
\ttfamily
#2
\end{tcolorbox}
}
\subsubsection{RQ1 Prompt}
\promptbox{Generalized Statement Execution Prediction Prompt}{
Here's some \{lang\} code. Each example highlights a single statement of (assignment, branch, or function calls) and shows you what the variable values look like just before it runs.

Your goal? Figure out what the result will be right after that statement runs.

Here are \{shot\} examples to walk you through it:
----------------------------------------
}

\promptbox{Assignment Prediction Prompt}{
You’re given some \{lang\} code and one specific assignment line.

Here are the local variables just before that line runs. Can you figure out what the value of the assignment will be afterwards?

Code Snippet:
\{lang\}
\{code\}

Statement: \{statement\}

Before Values:
\{variables\}

Answer using <ans></ans> tags, Do not include any extra information.
}

\promptbox{Boolean Prediction Prompt}{
Here’s a branch(if)/Boolean statement in \{lang\}, and the values of the variables it uses.

Will the branch run? Answer 'Yes' or 'No'.

Code:
\{lang\}
\{code\}

Branch Statement: \{statement\}

Condition Variables:
\{variables\}

Answer using <ans></ans> tags, Do not include any extra information.
}

\promptbox{Function Call Prompt}{
Here’s a function or API call in \{lang\} with some parameters.

Based on the inputs, what will it return?

Code:
\{lang\}
\{code\}

Call: \{statement\}

Parameter Values:
\{variables\}

Answer using <ans></ans> tags, Do not include any extra information.
}

\subsubsection{RQ2 Prompts}
\promptbox{Generalized Block Execution Prediction Prompt}{
Take a look at the \{lang\} code blocks. One statement is highlighted in each.

You’ll also see the input values going into the function. Based on those, try to figure out what the highlighted line will do.

Here are \{shot\} examples that show how it works:
----------------------------------------
}

\promptbox{Block Prediction Prompt}{
Here’s a full function in \{lang\} and a line of code inside it we care about.

Given the function’s inputs, what value will that line produce?

Code:
\{lang\}
\{code\}

Statement: \{statement\}

Inputs:
\{inputs\}

Answer using <ans></ans> tags
}

\promptbox{Generalized input\_output prompt}{
Here’s some \{lang\} code. You’ll either get the inputs or the outputs, but not both.

Your task is to fill in the missing part—predict the output if you know the input, or figure out what input must’ve produced the output.

Check out these \{shot\} examples for reference:
----------------------------------------
}

\promptbox{Output Prompt}{
Here’s some \{lang\} code and the inputs passed into it.

What output do you expect from it?

Code:
\{lang\}
\{code\}

Inputs:
\{input\}

Answer using <ans></ans> tags
}

\promptbox{Input Prompt}{
You know the output of a piece of \{lang\} code. Can you figure out what the input must’ve been?

Code:
\{lang\}
\{code\}

Output:
\{output\}

Answer using <ans></ans> tags
}

\subsubsection{RQ3 Prompts}
\promptbox{Generalized Loop Prediction Prompt}{
Let’s explore some loops in \{lang\}. You’ll get the full loop structure along with the input values used in the code.

I’ll ask you questions about how the loop body or post-loop values behave with those inputs.

Here’s how it works with \{shot\} example(s):
----------------------------------------
}
\promptbox{Iteration Prediction}{
Take a look at this {lang} loop with some given inputs.\\
Question:\{question\}\\
Code:\\
\{lang\}\\
\{code\}\\
Answer using <ans></ans> tags.
}

\promptbox{In-Loop Prediction}{
This is a \{lang\} loop and what the input to the function looks like.\\
I’ll ask you something about what happens inside the loop body.\\
Code:\\
\{lang\}\\
\{code\}\\
Question:\\
\{question\}\\
Answer using <ans></ans> tags
}

\promptbox{Post-Loop Prediction}{
This is a \{lang\} loop and what the input to the function looks like.\\
I’ll ask you something about what happens after the loop body.\\
Code:\\
\{lang\}\\
\{code\}\\
Question:\\
\{question\}\\
Answer using <ans></ans> tags
}

\promptbox{Branch Prediction Prompt}
{
Here’s a branch (if) block statement in \{lang\}.

Will the branch run given the function call? Answer 'Yes' or 'No'.

Code:
\{lang\}
\{code\}

Question: \{question\}

Answer using <ans></ans> tags, Do not include any extra information.
}
\promptbox{Alias Prediction}{
Here's some \{lang\} code with two pointer variables:

- Pointer A: '\{pointer\_1\}'

- Pointer B: '\{pointer\_2\}'

Do these pointers reference the same memory address? Answer "Yes" or "No".

Code:
\{lang\}
\{code\}

Function Input:
\{input\}

Question:
Do '\{pointer\_1\}' and '\{pointer\_2\}' in (line \{line\_1\}) point to the same memory location?

Put your answer in <ans></ans> tags.
}

\subsubsection{RQ4 Prompts}
\promptbox{Assignment CoT}{
Let’s figure out the result of the assignment: '\{statement\}'

You’ve got the current variable values: \{variables\}

Think through the right-hand side, then update the left-hand side with the result.
}

\promptbox{Boolean CoT}{
Here’s the condition: '\{statement\}'

These are the variable values: \{variables\}

Evaluate the condition. Is it true or false? That tells you if the branch runs.
}

\promptbox{Function Call CoT}{
This is the function call: '\{statement\}'

With these parameter values: \{variables\}

Figure out what the function does and predict the return value.
}

\promptbox{Block Prediction CoT}{
First, trace the execution flow till the highlighted statement \{statement\} and \{input\} of the given input,

Then identify the variables associated with the statement

Next, use the trace execution flow to evaluate the statement

What value does the statement produce?
}

\promptbox{Output Prediction CoT}{
We’re given inputs: \{input\}

Walk through the code step by step.

Watch how the values change until we get the final output.

Check that it matches what the function should return.
}

\promptbox{Input Prediction CoT}{
We know the output: \{output\}

Work backwards—what input could’ve led to that?

Figure out what had to happen in the code, and reverse it to get the input.
}

\promptbox{Iteration Prediction CoT}{
Start the loop using the initial values.

Check the condition, run the body, update, and repeat.

Keep going until the loop ends.
}

\promptbox{Loop in-Value CoT}{
Look at the variables at the start of this iteration.

Go through each line in the loop body.

What happens to the variables by the end?
}

\promptbox{Loop Post-Value CoT}{
See why the loop stopped (condition failed).

Check the final values of all changed variables.

What did the last iteration do before ending?

What would be the variable value after loop termination?
}

\promptbox{Overall Statement Prediction Prompt (1-shot) with CoT steps}{
Here's some Python code. Each example highlights a single statement of (assignment, branch, or function calls) and shows you what the variable values look like just before it runs.
Your goal? Figure out what the result will be right after that statement runs.\\

Here are 1 example to walk you through it:\\

----------------------------------------
EXAMPLE 1: ––––––––––––––––––––\\
Here’s a function or API call in Python with some parameters.\\
Based on the inputs/parameter values, what will it return?\\

\textbf{Code:}\\
Python \\
\{in-context Code\}\\

\textbf{Function Call:} \{statement with function call\}\\
\textbf{Parameter Values:}
\{values\}\\

Let's think step by step:\\
This is the function call: \{statement\}

With these parameter values: \{variables\}

Figure out what the function does and predict the return value.

\textbf{Therefore the final answer is:}<ans> \{Ground Truth\} </ans>\\

Now, please solve the following new problem.\\

You’re given some Python code and one specific assignment line.
Here are the local variables just before that line runs. Can you figure out what the value of the assignment will be afterwards?\\

\textbf{Code:}\\
Python\\
\{Query Code\}\\

\textbf{Statement:} \{selected statement\}\\

\textbf{Before Values:}
\{values\}\\

Answer using <ans></ans> tags, Do not include any extra information.\\
}
\scalebox{0.95}{
\begin{minipage}{1.05\textwidth}
\subsubsection{RQ5 Prompts}

\promptbox{Statement Prediction Prompt with Abstract Mapping}{
You’re given some Python code and one specific assignment line.\\
Here are the local variables just before that line runs. Can you figure out what the value of the assignment will be afterwards?\\

\textbf{Code:}\\
Python\\
\{Query Code\}\\

\textbf{Statement:} \{selected statement\}\\

\textbf{Before Values:}
\{values\}\\

You have to give your value prediction using the given quantization rules: \{rules\_list\}\\

Answer using <ans></ans> tags, Do not include any extra information.\\
}
\end{minipage}
}

% \scalebox{0.9}{
% \begin{minipage}{1.05\textwidth}

% \end{minipage}
% }

% \ray{in the appendix, it might be better to show each task similar to listing 1 and 2 of CruxEval}

\subsection{Additional Results}

\subsubsection{RQ1}
\begin{figure}[htbp]
    \centering
    \includegraphics[width=0.95\textwidth]{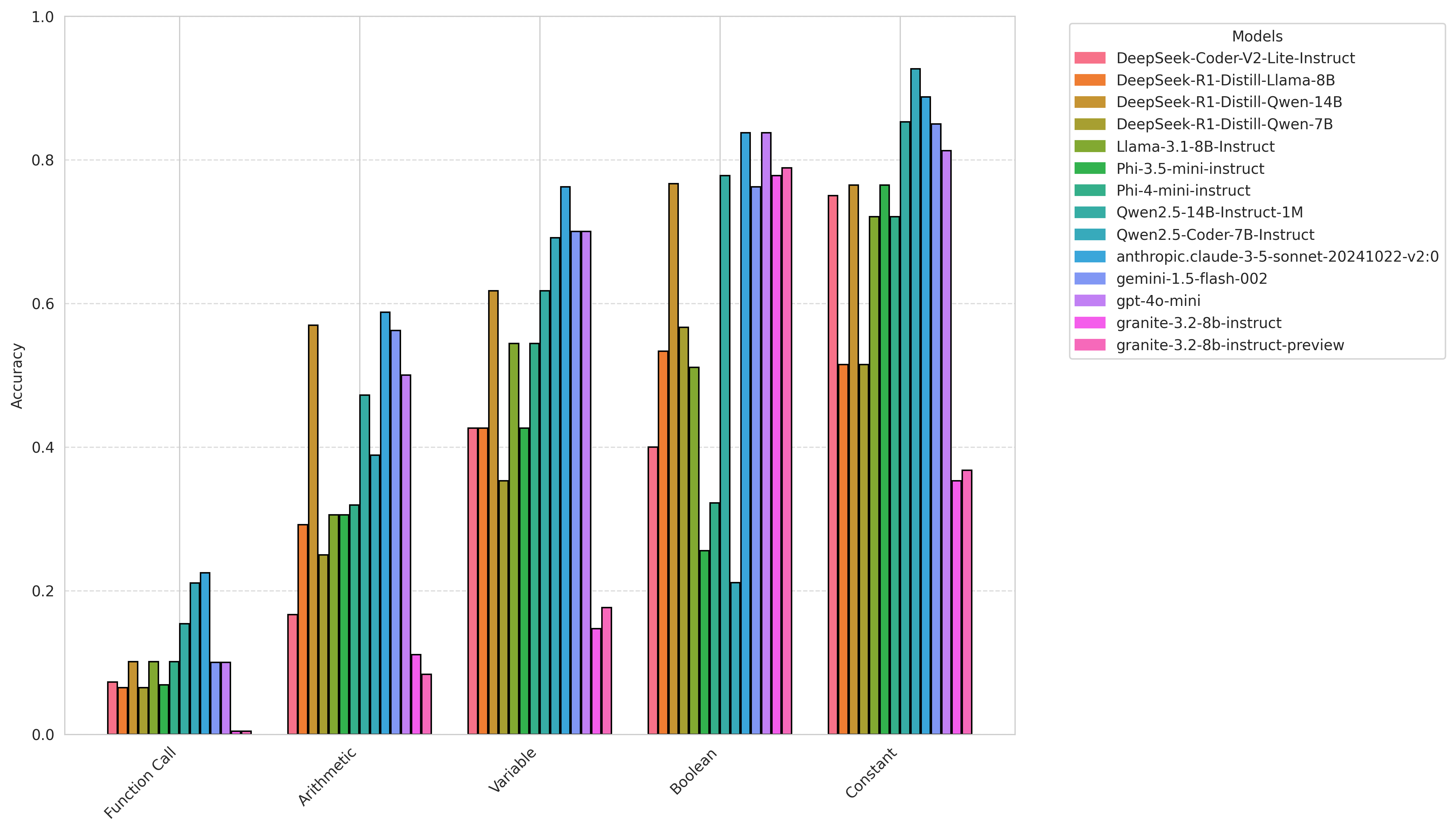}
    \caption{\textbf{RQ1} Statement type accuracy across Models (Python)}
    \label{app:rq1_image2}
\end{figure}

\begin{figure}[htbp]
    \centering
    \includegraphics[width=0.98\textwidth]{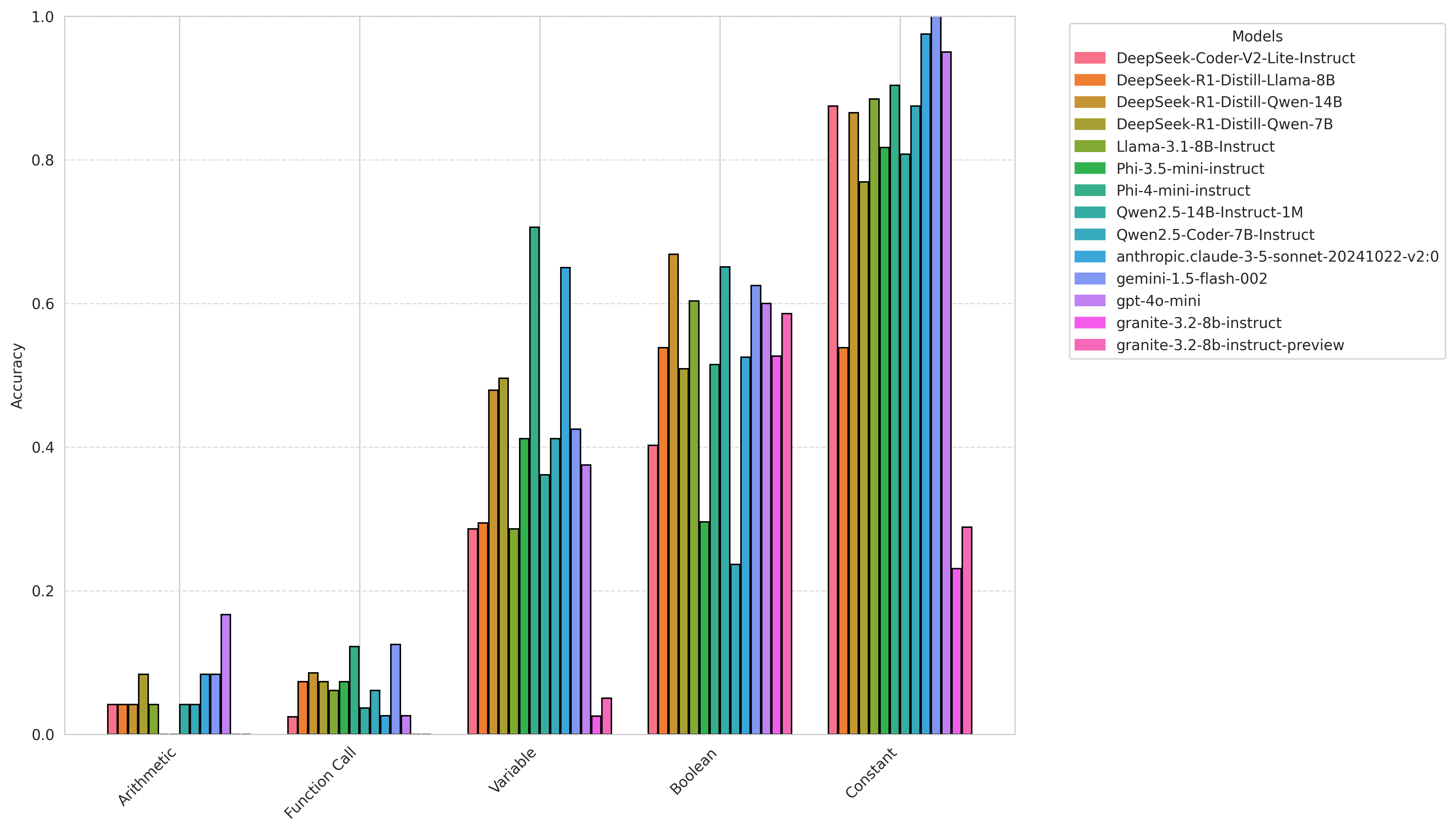}
    \caption{\textbf{RQ1} Statement type accuracy across Models (C)}
    \label{app:rq1_image4}
\end{figure}

\begin{figure}[htbp]
    \centering
    \includegraphics[width=0.98\textwidth]{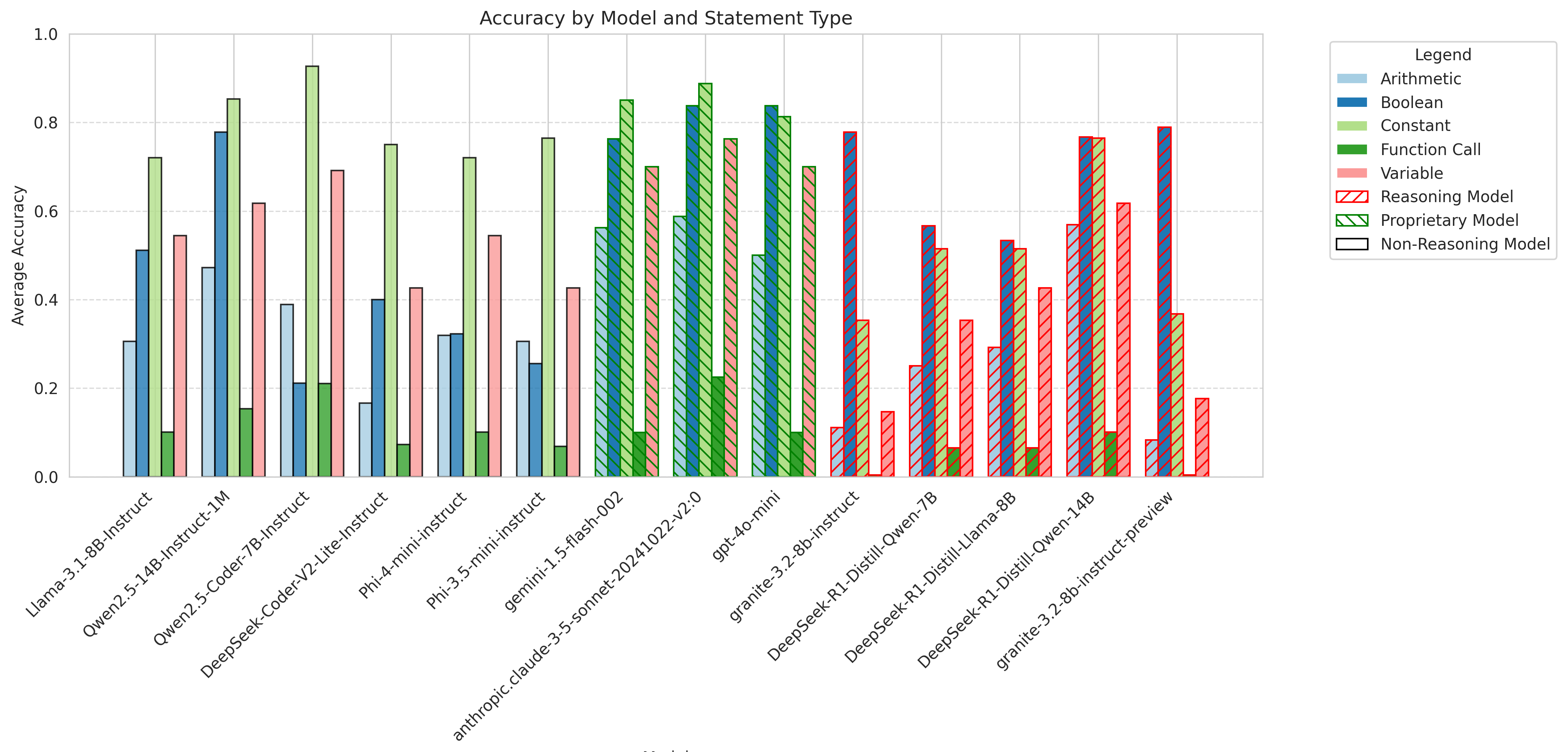}
    \caption{\textbf{RQ1} Statement type accuracy across Models (Python)}
    \label{app:rq1_image1}
\end{figure}

\begin{figure}[htbp]
    \centering
    \includegraphics[width=0.95\textwidth]{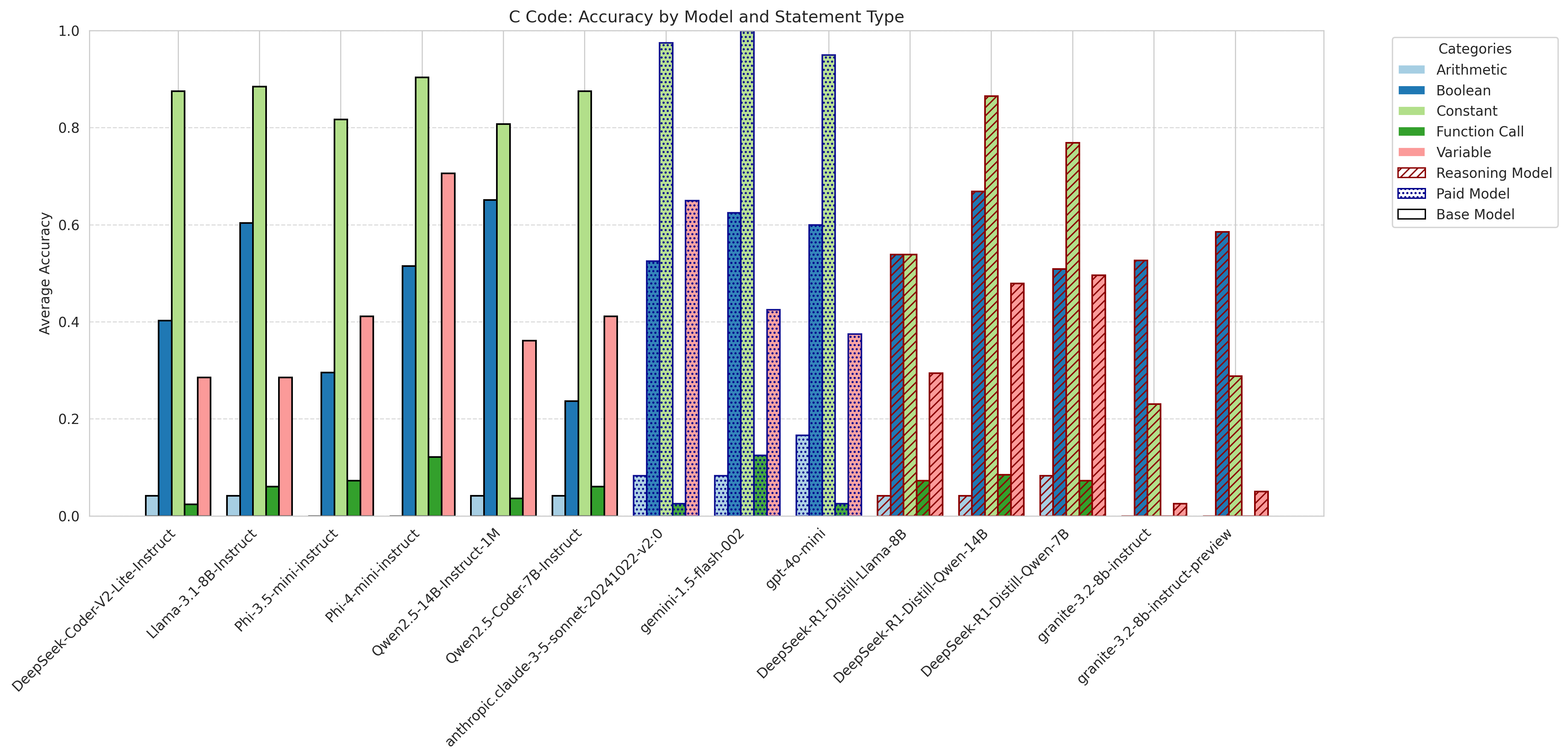}
    \caption{\textbf{RQ1} Statement type accuracy across Models (C)}
    \label{app:rq1_image3}
\end{figure}

\figref{app:rq1_image2} and \figref{app:rq1_image4} depict each model's capability on individual statement types. 
\figref{app:rq1_image1} and \figref{app:rq1_image3} show the performance of all the models across five types of statements for languages Python and C.

\subsubsection{RQ4}
\begin{figure}[htbp]
    \centering
    \includegraphics[width=0.95\textwidth]{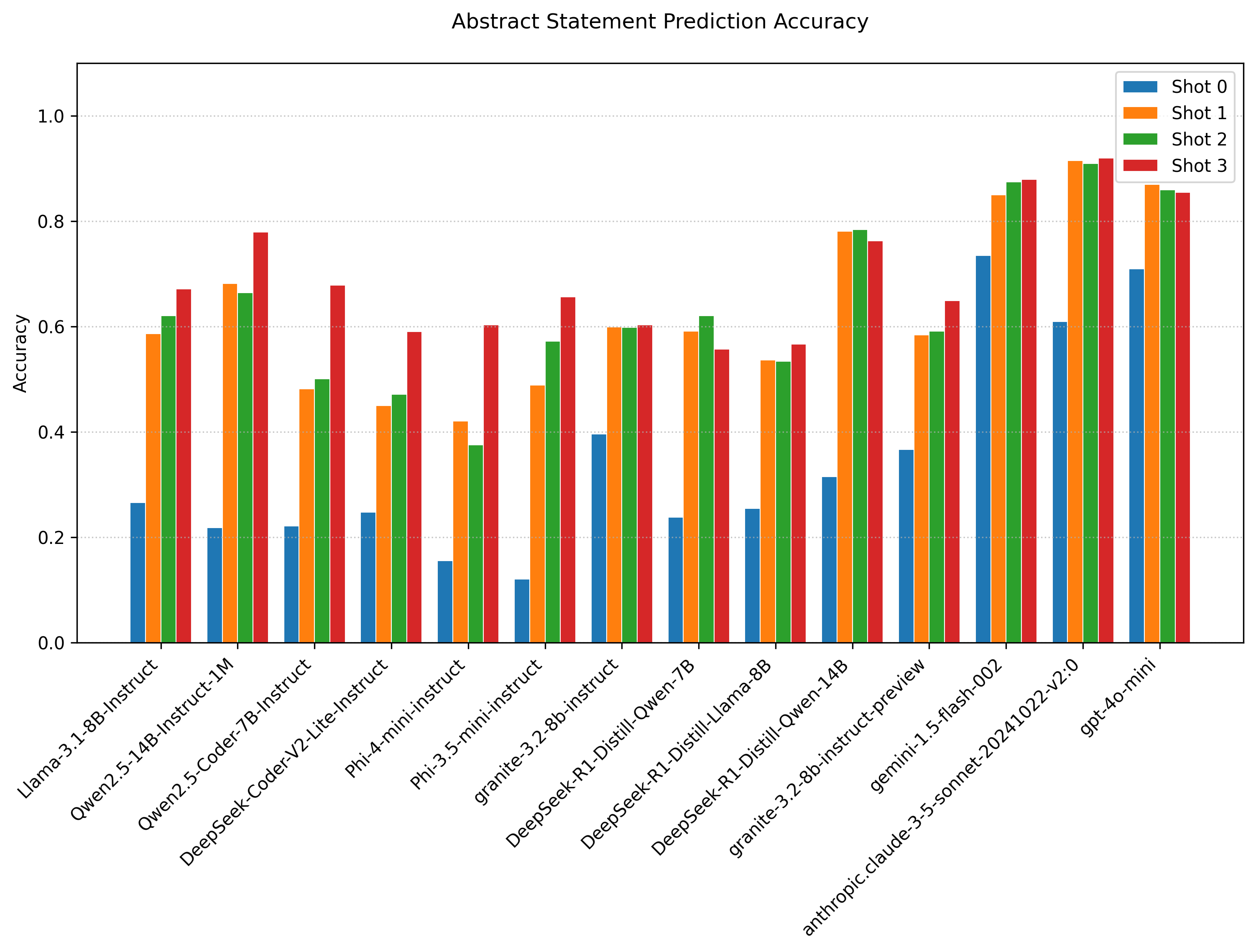}
    \caption{\textbf{RQ4} Models' Performance with increasing shots from 0 to 3 (Abstract Value Prediction).}
    \label{app:rq4_image1}
\end{figure}

\begin{figure}[htbp]
    \centering
    \includegraphics[width=0.95\textwidth]{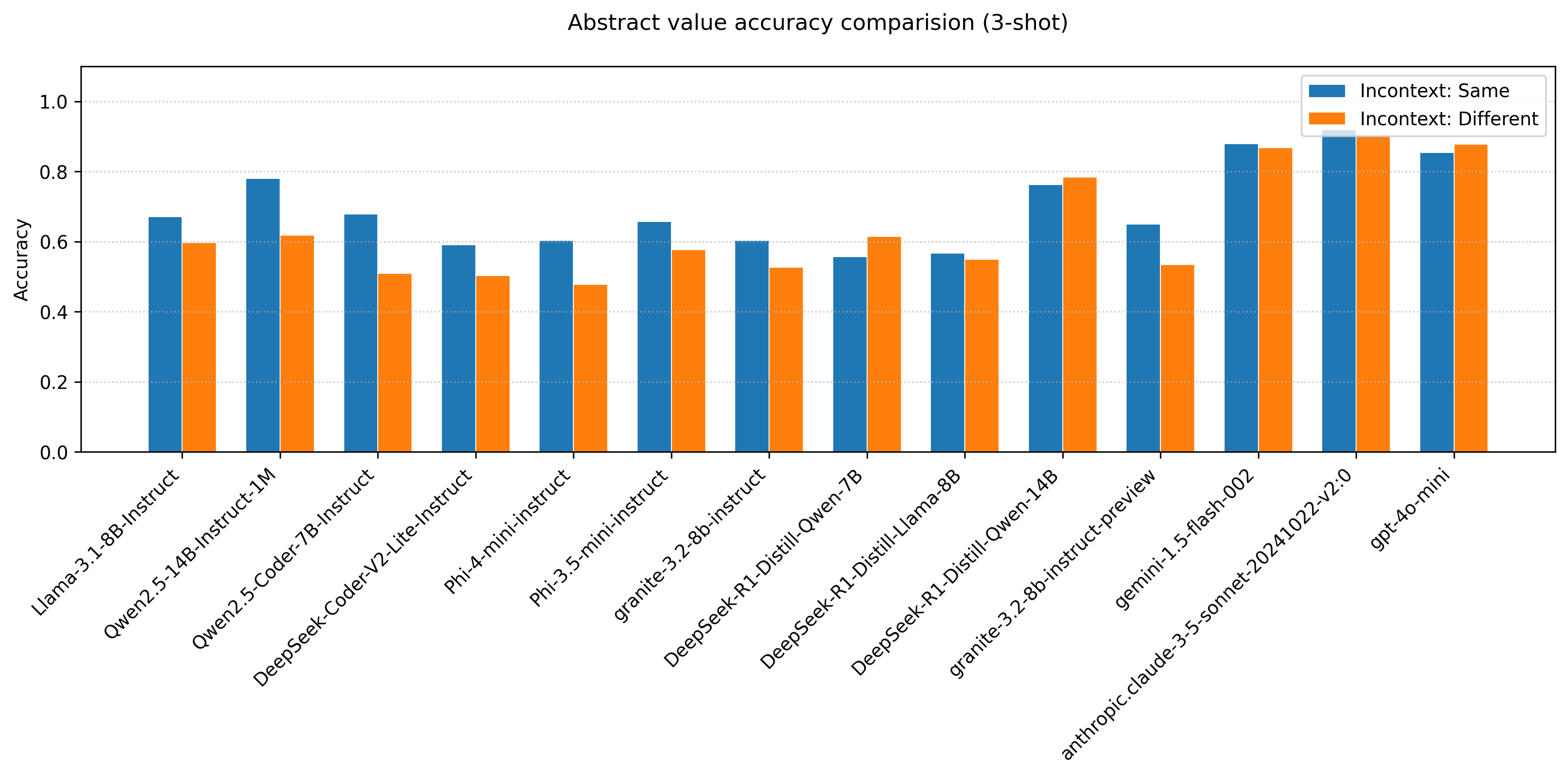}
    \caption{\textbf{RQ4} Models' Performance with random and same function in-context Examples.}
    \label{app:rq4_image2}
\end{figure}

\begin{figure}[htbp]
    \centering
\includegraphics[width=0.95\textwidth]{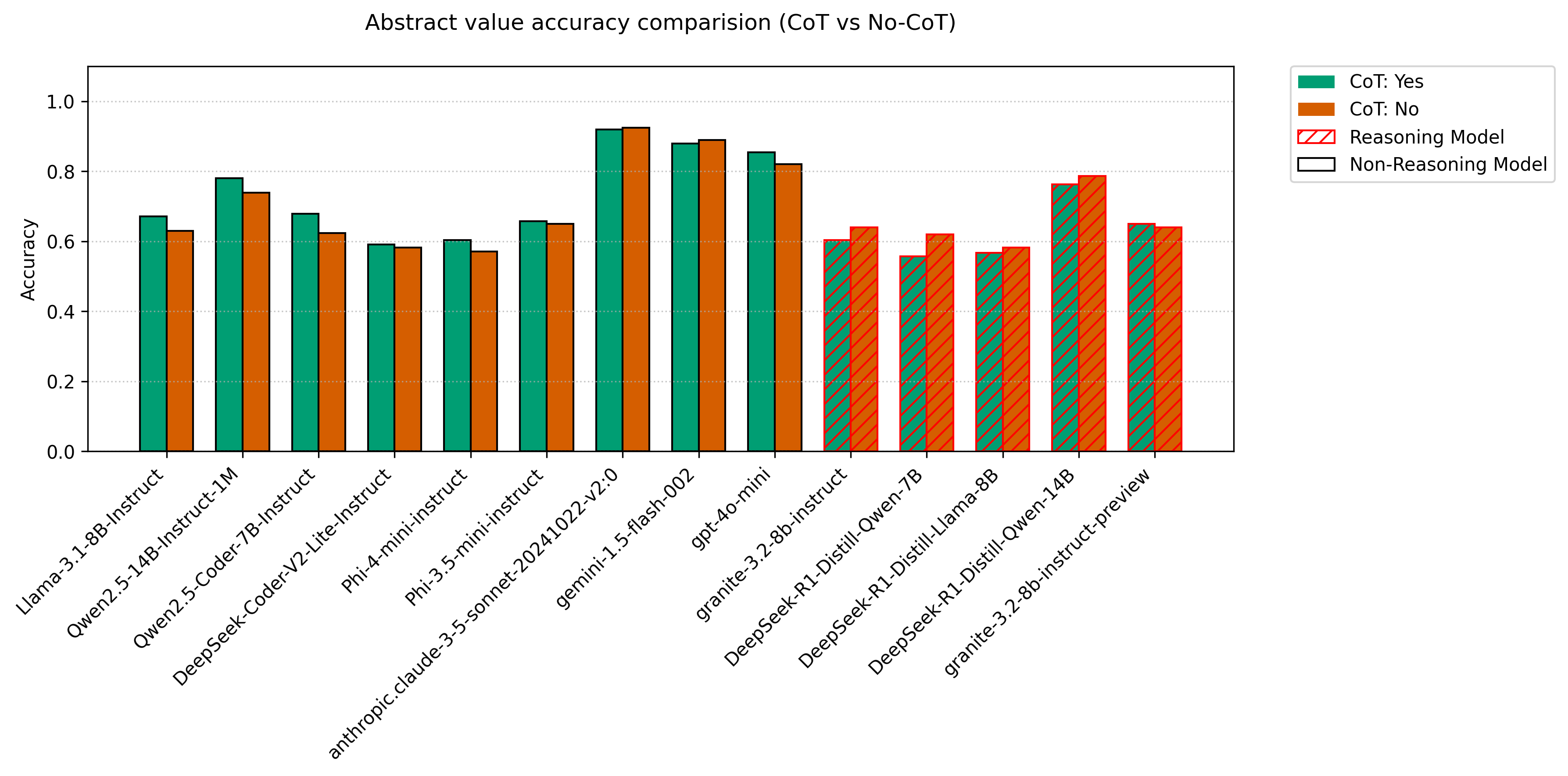}
    \caption{\textbf{RQ4} Models' Performance CoT vs No-CoT}
    \label{app:rq4_image3}
\end{figure}

In \figref{app:rq4_image1}, we show that for most of the models, adding the number of shots/in-context examples helps the models. \figref{app:rq4_image2} demonstrates that selecting in-context examples in a more controlled way, for example, selecting the same function as in-context examples, helps the models reason better. Finally, \figref{app:rq4_image3} shows whether adding a Chain of Thought (CoT) with the in-context examples can help improve the performance. 

\subsubsection{RQ5} \label{app:RQ5 results}
\begin{figure}[H]
    \centering
\includegraphics[width=0.95\textwidth]{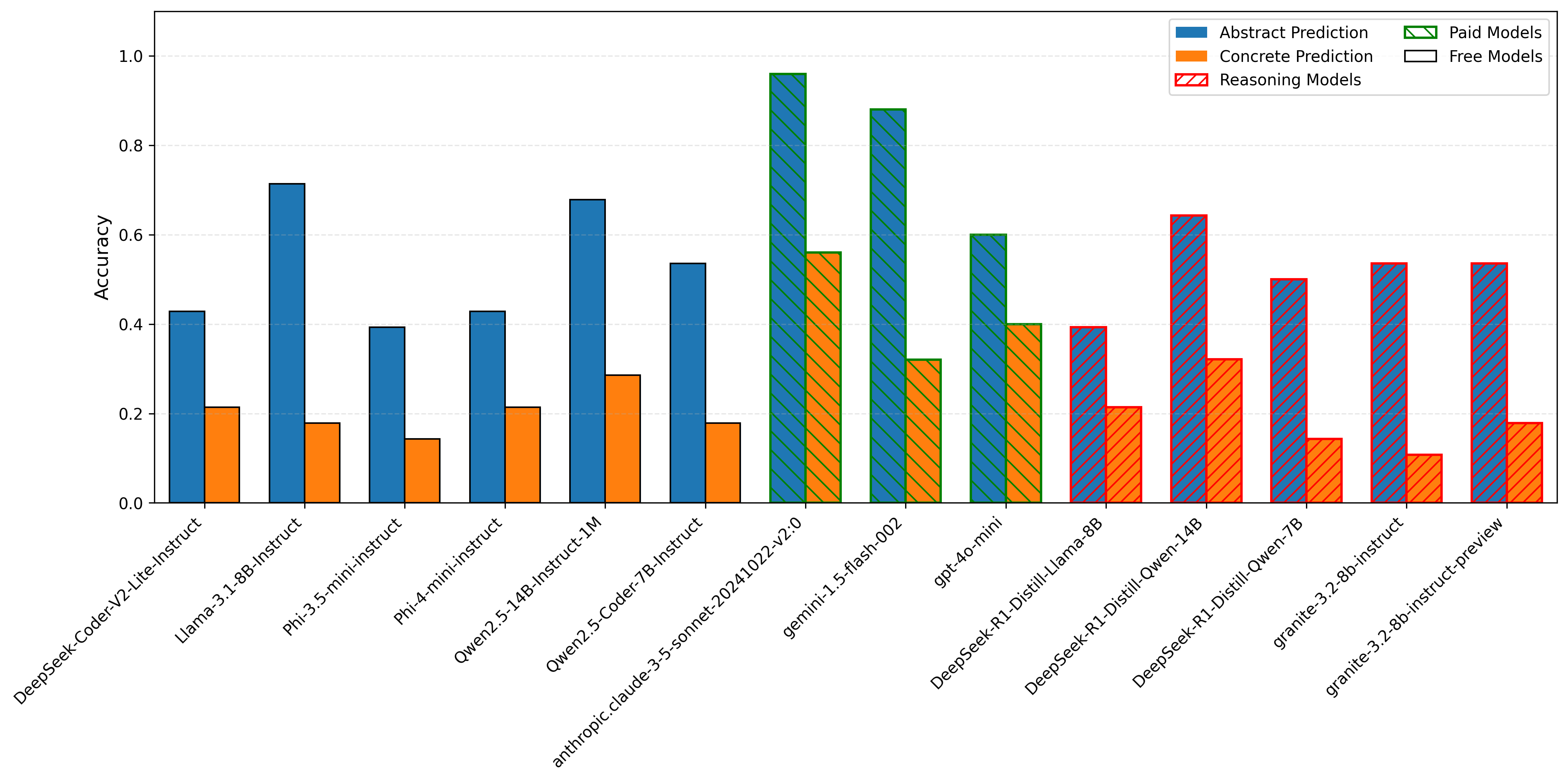}
    \caption{\textbf{RQ5:} Post-Loop value prediction abstract vs concrete values (3-shots)}
    \label{app:rq5_image1}
\end{figure}

In \figref{app:rq5_image1}, we compare abstract value  vs concrete value prediction for post-loop values. Though the models struggle with concrete value prediction, they can improve the performance for predicting the range/approximation of the concrete value.

% \begin{figure}[htbp]
%     \centering
%     \begin{subfigure}[b]{0.48\textwidth} 
%         \includegraphics[width=\textwidth]{RQ3_Results/after_comparison.png}
%      %   \caption{Python}
%     \end{subfigure}
%     \hfill

%     \caption{\textbf{RQ5:} Can models reason about an approximation of code semantics? }
%     \label{app:rq5_image1}
% \end{figure}

\begin{table}[htbp]
\centering
\caption{Model Performance Comparison on Abstract Value Prediction} \label{tab:random baseline}
\begin{tabular}{lcccc}
\hline
Model Name & Random Baseline & 0-shot & 3-shot \\
\hline
DeepSeek-Coder-V2 & 0.084 & 0.247 & 0.504 \\
DeepSeek-R1-Distill-Llama & 0.112 & 0.255 & 0.549 \\
DeepSeek-R1-Distill-Qwen-14B & 0.169 & 0.316 & 0.784 \\
DeepSeek-R1-Distill-Qwen-7B & 0.113 & 0.238 & 0.614 \\
Llama-3.1-8B & 0.104 & 0.266 & 0.597 \\
Phi-3.5-mini & 0.077 & 0.121 & 0.577 \\
Phi-4-mini & 0.088 & 0.156 & 0.478 \\
Qwen2.5-14B & 0.176 & 0.218 & 0.618 \\
Qwen2.5-Coder-7B & 0.099 & 0.222 & 0.509 \\
granite-3.2-8b & 0.152 & 0.396 & 0.528 \\
granite-3.2-8b & 0.147 & 0.367 & 0.535 \\
\hline
\end{tabular}
\end{table}

% ==== APPENDIX A.8 API Definition Ablation Study ====

\subsection{API Definition Ablation Study} \label{app:api}
In Task 2, when predicting the output for output of an API call, we conducted additional experiments to evaluate whether providing API definitions improves model performance. We tested two types of settings: (i) with No API definition and (ii) with API definitions.

We ran our evaluation on the open-source models and evaluated all the 248 function call prediction examples from our dataset. Table~\ref{tab:api-ablation} shows the results on the best-performing open-source models:

\begin{table}[h]
\centering
\caption{Accuracy of API prediction with different API definition strategies}
\label{tab:api-ablation}
\begin{tabular}{lcc}
\hline
Model & No API definitions & API implementation \\
\hline
Qwen 2.5-7B & 0.206 & 0.226 \\
Qwen 2.5-14B & 0.182 & 0.194 \\
Phi-4 & 0.125 & 0.089 \\
Llama-3.1-8B & 0.105 & 0.089 \\
\hline
\end{tabular}
\end{table}

The results indicate that providing API definitions does not significantly improve performance, and in some cases slightly degrades it. This supports our hypothesis that the fundamental limitation for fine-grained code reasoning is not lack of API knowledge, but rather the models' inability to reason about statement-level and block-level semantics. 

\subsection{Comparison with Existing Benchmarks}
\label{app:benchmark-comparison}
We did a partial evaluation of the output prediction task on our evaluation framework on the three best-performing open-source models on CodeSense, and we present the results in Table \ref{tab:cruxeval-codesense-comparison}. Our result shows that the models perform significantly worse in CodeSense than CruxEval. 

\begin{table}[H]
\centering
\caption{Output prediction accuracy comparison: CruxEval vs. CodeSense}
\label{tab:cruxeval-codesense-comparison}
\begin{tabular}{lccc}
\toprule
\textbf{Model} & \textbf{CruxEval} & \textbf{CodeSense} & \textbf{Drop} \\
\midrule
DeepSeek-R1-Distill-Qwen-14B & 0.75 & 0.37 & 0.38 \\
Qwen2.5-14B & 0.52 & 0.27 & 0.25 \\
Qwen2.5-Coder-7B & 0.50 & 0.30 & 0.20 \\
\bottomrule
\end{tabular}
\end{table}

\subsection{Variance Analysis}
\label{sec:variance}

We have run the statement prediction task, and the results in Table \ref{tab:variance} across three runs demonstrate that the variance across multiple runs is minimal, and this doesn’t change our core findings.  

\begin{table}[H]
\centering
\caption{Variance analysis across three runs on statement prediction task}

\label{tab:variance}
\begin{tabular}{lcccc}
\toprule
\textbf{Model} & \textbf{Run 1} & \textbf{Run 2} & \textbf{Run 3} & \textbf{Mean ± Std Dev} \\
\midrule
Qwen2.5-14B & 44.4\% & 44.4\% & 43.3\% & 44.0\% ± 0.6\% \\
DeepSeek-R1-Distill-Qwen-14B & 42.0\% & 41.8\% & 43.8\% & 42.5\% ± 1.0\% \\
Qwen2.5-Coder-7B & 38.4\% & 38.4\% & 38.4\% & 38.3\% ± 0.0\% \\
DeepSeek-R1-Distill-Llama-8B & 26.4\% & 25.7\% & 27.3\% & 26.5\% ± 0.8\% \\
\bottomrule
\end{tabular}
\end{table}

\subsection{Different Prompting techniques for Statement Prediction} 
\label{sec:same}

Table \ref{tab:statement-prediction-type} shows the difference between two prompting strategies: using in-context examples of the \textit{same} statement type as the query versus examples of a \textit{different} statement type.

\begin{table}[H]
\centering
\caption{Statement Prediction Performance by Type}
\label{tab:statement-prediction-type}
\begin{tabular}{lcc}
\toprule
\textbf{Model} & \textbf{Same Type Statement} & \textbf{Different Type Statement} \\
\midrule
Qwen2.5-14B-Instruct-1M & 0.44 & 0.42 \\
DeepSeek-R1-Distill-Qwen-14B & 0.42 & 0.39 \\
Qwen2.5-Coder-7B-Instruct & 0.38 & 0.37 \\
Llama-3.1-8B-Instruct & 0.32 & 0.29 \\
Phi-4-mini-instruct & 0.30 & 0.25 \\
\bottomrule
\end{tabular}
\end{table}

\subsection{Function Size Analysis}
\label{sec:function-size}

Table \ref{tab:function-size-categories} shows how we categorise function difficulties based on lines of code.
\label{tab:Function}
\begin{table}[H]
\centering
\caption{Function Size Categories}
\label{tab:function-size-categories}
\begin{tabular}{lc}
\toprule
\textbf{Category} & \textbf{Length (Lines of Code)} \\
\midrule
Small & $\text{length} \leq 9$ \\
Medium & $10 < \text{length} \leq 19$ \\
Large & $\text{length} \geq 20$ \\
\bottomrule
\end{tabular}
\end{table}

\subsection{Examples of Fine-Grained Insights}
\label{sec:fine-grained-insights}

Codesense’s fine-grained task can uncover reasoning failures on code semantics that are not visible to the coarse-grained benchmarks. For example, consider the simple function from our benchmark

\begin{verbatim}
def _is_ascii(s):
    if isinstance(s, str):
        for c in s:
            if ord(c) > 255:
                return False
        return True
    return _supports_unicode(s)

_is_ascii(' 123456789#')
\end{verbatim}

\noindent \textbf{Question:} ``How many times will the loop on line 3 iterate?''\\
\textbf{Ground Truth:} 11 (The length of the input string \verb|s|, which contains a leading space, digits 1–9, and the \verb|#| character).

However, models such as \texttt{Qwen2.5-Coder-7B} incorrectly respond with \textbf{10}. This error reveals that the model fails at a fundamental level: it cannot correctly reason about string iteration, specifically miscounting the characters in a simple string literal.

%%%%%%%%%%%%%%%%%%%%%%%%%%%%%%%%%%%%%%%%%%%%%%%%%%%%%%%%%%%%

\end{document}